\documentclass[twocolumn,superscriptaddress,aps,prd,amsmath,amsfonts,nofootinbib]{revtex4-1}

\newcommand{\ud}{\mathrm{d}}

\newcommand{\half}{{\tfrac{1}{2}}}

\newcommand{\bs}{\begin{split}}
\newcommand{\es}{\end{split}}

\newcommand{\muh}{{\hat{\mu}}}
\newcommand{\nuh}{{\hat{\nu}}}

\newcommand{\ob}{{\bar{\omega}}}
\newcommand{\vb}{{\bar{v}}}
\newcommand{\alphabar}{{\bar{\alpha}}}

\usepackage{verbatim}
\usepackage{float}
\usepackage{color}
\usepackage[hyperindex,breaklinks]{hyperref}
\hypersetup{
    pdfborder={0 0 0.5 0},
    linkbordercolor={0.6 1 0.6},
    citebordercolor={0.6 1 0.6},
    urlbordercolor={0.6 1 0.6},
    pdftitle={Casimir Energy of a Long Wormhole Throat},
    pdfauthor={Luke M. Butcher}}
\usepackage{slashed}
\usepackage{graphicx}
\usepackage{subfigure}
\usepackage{psfrag}
\usepackage{mathtools}
\usepackage{amsthm}
\theoremstyle{plain}

\begin{document}

\title{Casimir Energy of a Long Wormhole Throat}
\author{Luke M. Butcher}
\email[]{l.butcher@mrao.cam.ac.uk}
\affiliation{Astrophysics Group, Cavendish Laboratory, University of Cambridge, J J Thomson Avenue, Cambridge, CB3 0HE, UK}
\affiliation{Kavli Institute for Cosmology Cambridge, Madingley Road, Cambridge, CB3 0HA, UK}
\date{May 6, 2014}

\begin{abstract}
We calculate the Casimir energy-momentum tensor induced in a scalar field by a macroscopic ultrastatic spherically-symmetric long-throated traversable wormhole, and examine whether this exotic matter is sufficient to stabilise the wormhole itself. The Casimir energy-momentum tensor is obtained (within the $\mathbb{R}\times S_2$ throat) by a mode sum approach, using a sharp energy cut-off and the Abel-Plana formula; Lorentz invariance is then restored by use of a Pauli-Villars regulator. The massless conformally-coupled case is found to have a logarithmic divergence (which we renormalise) and a conformal anomaly, the thermodynamic relevance of which is discussed. Provided the throat radius is above some fixed length, the renormalised Casimir energy-density is seen to be negative by all timelike observers, and almost all null rays; furthermore, it has sufficient magnitude to stabilise a long-throated wormhole far larger than the Planck scale, at least in principle. Unfortunately, the renormalised Casimir energy-density is zero for null rays directed exactly parallel to the throat, and this shortfall prevents us from stabilising the ultrastatic spherically-symmetric wormhole considered here. Nonetheless, the negative Casimir energy does allow the wormhole to collapse extremely slowly, its lifetime growing without bound as the throat-length is increased. We find that the throat closes slowly enough that its central region can be safely traversed by a pulse of light.
\end{abstract}
\maketitle

\section{Introduction}\label{intro}
The idea of a ``bridge'' of curved space, linking two otherwise distant regions, has served as a rich basis for thought-experiments, and a valuable test-bed for questions at the interface of gravitational and quantum theory. These \emph{wormholes} have found varied applications, from models of fundamental particles \cite{Misner75}, to ingredients of a mechanism that would supposedly suppress the cosmological constant \cite{Coleman88,Klebanov89}. More recently, a fascinating connection between wormholes and quantum entanglement has been conjectured \cite{Raamsdonk10,Maldacena13} which has played a key role in the ongoing debate over the existence of a ``firewall'' behind a black hole event horizon \cite{Almheiri13}. Lastly, and most provocatively of all, there is the question of whether stable \emph{traversable} wormholes can exist, and if so, whether anything prevents their being used as time-machines \cite{Morris88,Visser90, Novikov90}.

In this paper we will focus on traversable wormholes, and explore a mechanism which might allow them to exist, at least in principle. As is well known, the key impediment to their stability is the need for exotic matter: negative energy is required, as averaged along a null geodesic that threads the throat and escapes to infinity \cite{Morris88}. The only experimentally verified phenomenon expected to produce negative energy is the Casimir effect \cite{Casimir48}, wherein conductive plates are introduced to empty space, and these plates impose boundary conditions on the vacuum state of a quantum field. In many cases, the new ground state energy is less than that of the original (zero-energy) vacuum, leading to the conclusion that a negative energy has been achieved. Adapting this phenomenon to the problem at hand, one would hope to induce a negative energy vacuum in the throat of a wormhole, presumably by capping its mouths with conductive plates (as in \cite{Khabibullin06}, for example). However, the plates themselves will inevitably possess some mass, and under reasonable assumptions\footnote{The plates should have a mass-to-charge ratio no less than the electron, and should be further apart than the electron's Compton wavelength \cite{Morris88}.} this positive energy will outweigh the negative energy between the plates when averaged along a null geodesic that escapes to infinity.

Fortunately, there remains a plausible route around this obstacle, which we shall presently explore. The idea is this: discard the conductive plates altogether, and ask whether the wormhole \emph{itself}, by virtue of its curvature and topology, can generate the negative Casimir energy it requires.

Now, cursory dimensional analysis would suggest that this mechanism can only stabilise a Planckian wormhole,\footnote{If it were possible to describe the wormhole/field system by a single characteristic length (the ``size'' of the wormhole) then it follows that $(\text{size})\sim (\text{Planck length})$ as there is no other quantity available with the correct dimensions. Of course, the wormhole/field system need not be characterised by a single length.} in which case the semiclassical approach (quantum field propagating on classical spacetime) would be expected to break down anyway.\footnote{This is the main criticism one can levy at the self-sustaining wormhole solution obtained in \cite{Hochberg97} by numerical techniques. Besides the presence of Planck-scale structure, this solution is also asymptotically ill-behaved: the time-directed killing vector diverges.} To avoid this pitfall, it is therefore desirable to optimise the shape of the wormhole so as to (a) \emph{increase} the magnitude of the (negative) Casimir energy-density it \emph{generates}, and (b) \emph{decrease} the magnitude of the negative energy-density it \emph{requires}. 

\begin{figure}
\includegraphics[scale=.45]{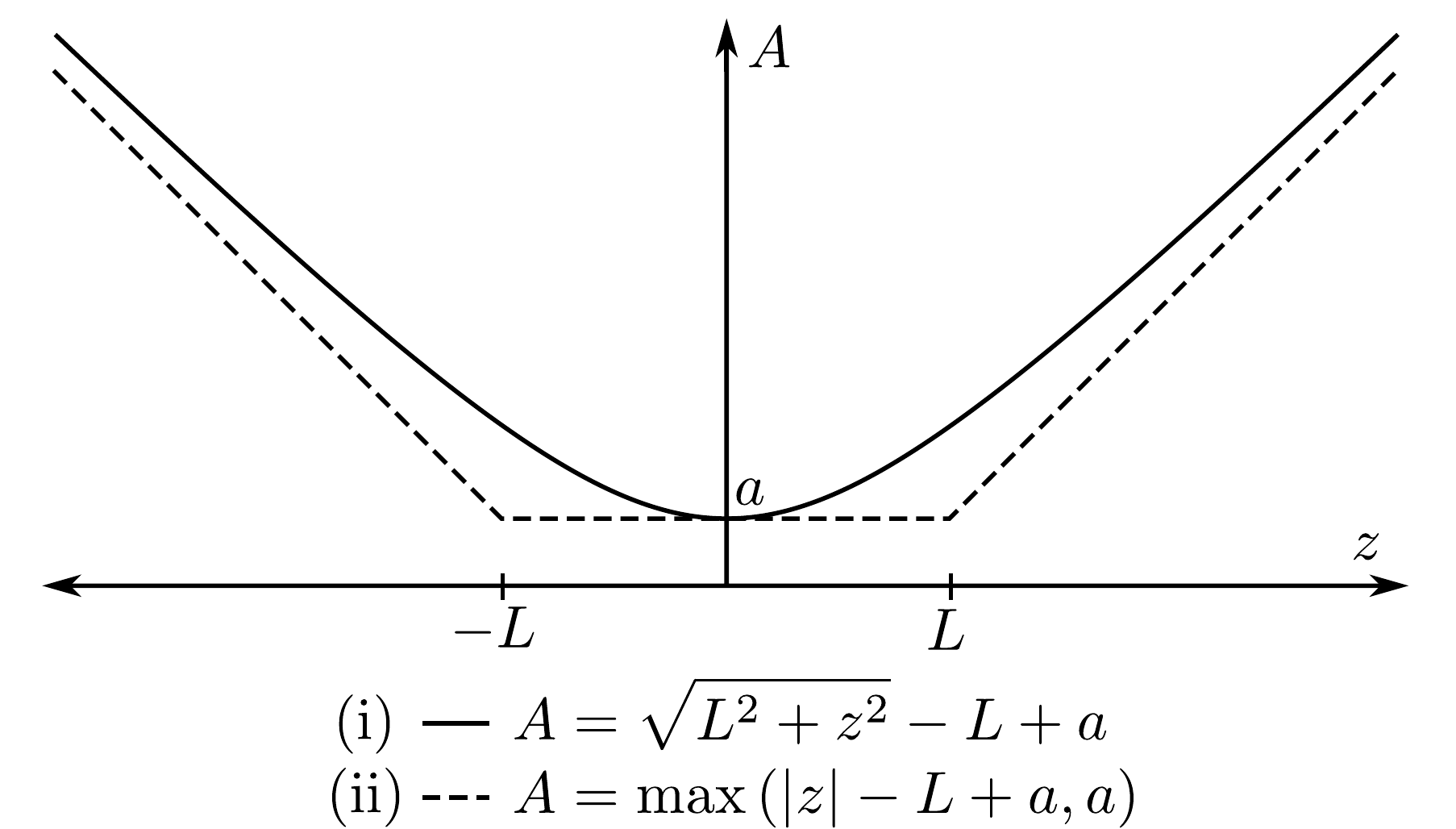}
\caption{(i) Spatial profile of the two-parameter wormhole (\ref{2par}); (ii) Spatial profile of a simple surgically-constructed wormhole with throat length $2L$ and throat radius $a$. }\label{2parprofile}
\centering
\end{figure}

One very simple way of achieving this is to make the wormhole much longer than it is wide. For the purpose of explaining this claim, let us consider the following spherically symmetric static traversable wormhole:\footnote{We set $c=1$, write $\kappa \equiv 8\pi G$, and adopt the sign conventions of Wald \cite{Wald}: $\eta_{\mu\nu}\equiv \mathrm{diag}(-1,1,1,1)$, $[\nabla_\mu,\nabla_\nu]v^\alpha\equiv R^{\alpha}_{\phantom{\alpha}\beta\mu\nu}v^\beta$, and $R_{\mu\nu}\equiv R^{\alpha}_{\phantom{\alpha}\mu\alpha\nu}$.}
\begin{align}\label{2par}\bs
\ud s^2 &= - \ud t^2 + \ud z^2 + A^2\left( \ud\theta ^2 + \sin^2\theta \ud \phi^2 \right),
\\ A&\equiv \sqrt{L^2 + z^2} - L + a.\es
\end{align}
As figure \ref{2parprofile} illustrates, this spacetime is a smooth realisation of a simple surgically-constructed wormhole which connects two flat regions with a spherically symmetric throat of length $2L$ and constant radius $a$.

The Einstein tensor for the spacetime (\ref{2par}) is straightforward to calculate, and reveals the energy-momentum tensor required by the wormhole:
\begin{align}\nonumber
T_{\muh\nuh}& = G_{\muh\nuh}/\kappa \\\nonumber
&= \frac{L^2}{(L^2 +z^2) A^2 \kappa}\mathrm{diag}\left(1,-1,\frac{A}{\sqrt{L^2+z^2}},\frac{A}{\sqrt{L^2+z^2}}\right)\\\label{2parT}
&\quad\,+ \frac{2L^2}{(L^2 +z^2)^{3/2} A \kappa}\mathrm{diag}\left(-1,0,0,0\right),
\end{align}
where the hats over indices indicate that components have been expressed in the \emph{orthonormal} basis along the $\{t,z,\theta,\phi\}$ coordinate lines. Let us assume $L \ge a$, and hence $A/\sqrt{L^2 + z^2}\le 1$. Consequently, the first tensor on the right-hand side of (\ref{2parT}) obeys all four energy conditions (null, weak, strong and dominant)\footnote{\label{nonexotic}It is a simple matter to prove that an energy-momentum tensor $T_{\muh\nuh} = \mathrm{diag}(1,-1,p,p)$ obeys the null, weak and dominant energy conditions if and only if $|p|\le 1$. Furthermore, the strong energy condition is obeyed if and only if $p \ge 0$. Consequently, this energy-momentum tensor (and any positive multiple thereof) will obey all the energy conditions if and only if $0\le p \le 1$.} and the second tensor can be interpreted as the exotic matter required to stabilise the wormhole. Note that the magnitude of this second tensor is greatest at $z=0$, where it takes the value $2/La\kappa$. This will serve as an adequate measure of the negative energy-density \emph{required} by the wormhole (\ref{2par}).

Now we turn to the Casimir energy-density \emph{generated} by the spacetime (\ref{2par}).  Near the centre of the wormhole, where the negative energy-density requirements are greatest, the throat radius is
\begin{align}
A = a + \frac{z^2}{2L} + O(z^4/L^3),
\end{align}
and on scales much smaller than the throat length, $A$ takes the constant value $a$ to a good approximation: $|\ud A/\ud z| \approx |z|/L \ll 1$. Hence, near the centre of the wormhole, quantum field modes with wavelengths much smaller than $L$ may as well be propagating in a throat of constant radius $a$, and can be expected to produce a Casimir energy-density of order $\hbar/a^4$. Clearly, if we hold $a$ constant and increase $L$, then (i) this approximation will improve, with the Casimir energy-density tending to a fixed value $O(\hbar/a^4)$, and (ii) the negative energy-density $O(1/La\kappa)$ required by the wormhole will decrease in magnitude. Ignoring the nonexotic matter, then, the Einstein equations (\ref{2parT}) take the form $\hbar/a^4 \sim 1/ L a\kappa$, from which it follows that
\begin{align}
a^2 \sim (l_p)^2 (L/a),
\end{align}
where $l_p$ is the Planck length. This suggests that $a$ and $L$ can \emph{both} be much larger than the Planck length, provided $L\gg a$.

What remains is to actually \emph{calculate} the Casimir energy-momentum tensor, and to check it possesses the required structure (in particular, \emph{negative} energy-density) to allow this rough argument to carry through. To simplify the calculation, we shall focus on the limit $L\to \infty$, in which the spacetime (\ref{2par}) becomes
\begin{align}\label{throat}
\ud s^2 &= - \ud t^2 + \ud z^2 + a^2\left( \ud\theta ^2 + \sin^2\theta \ud \phi^2 \right),
\end{align}
and the Ricci tensor is
\begin{align}\label{Ricci}
R_{\muh\nuh}&=a^{-2}\mathrm{diag}(0,0,1,1).
\end{align}
This will provide us with a good approximation to the Casimir energy-momentum generated by a wormhole with $L\gg a$, at least in the vicinity of the centre-point $z=0$. Calculating the Casimir energy-momentum tensor induced by (\ref{throat}) will be the main task this paper,\footnote{Note that because curvature coordinates are degenerate in the infinite throat (\ref{throat}) the treatment of vacuum energies by Anderson et.\ al.\ \cite{Anderson95} cannot be applied.} followed by an assessment of wormhole stability in section \ref{Stab}.

\section{Casimir Effect}\label{CEMT}
When one naively calculates the vacuum energy of a quantum field, one finds that it is infinite. The canonical remedy for this is \emph{normal ordering}, which subtracts this infinite constant and essentially \emph{defines} the vacuum energy to be zero. However, the ground state of a quantum field is dependent on the field's environment: the presence of conductive surfaces will impose boundary conditions, spacetime curvature will alter the field equations, and nontrivial topology will introduce additional constraints. Hence, even after one has fixed the vacuum energy of empty Minkowski spacetime at zero, energy \emph{differences} persist between vacuum states of different environments, and one is forced to admit that the vacua of nontrivial environments have nonzero energy.

Let us formulate this symbolically for the case at hand: let $\left|0\right>$ be the vacuum state of a quantum field $\varphi$ in the infinite throat spacetime (\ref{throat}), and let $\left|0_M\right>$ be the vacuum state of the same quantum field in empty Minkowski spacetime
\begin{align}\label{mink}
\ud s^2 = - \ud t^2 + \delta_{ij}\ud x^i \ud x^j;
\end{align}
then the Casimir energy-momentum tensor of $\varphi$ in the throat is 
\begin{align}\label{CasTdef}
T^{\mathrm{Casimir}}_{\mu\nu }\equiv \left<0\right| \widehat{T}_{\mu\nu}\left|0\right>-\left<0_M\right| \widehat{T}_{\mu\nu}\left|0_M\right>, 
\end{align}
that is, $T^{\mathrm{Casimir}}_{\mu\nu}$ is the vacuum energy-momentum that remains once we have accounted for the spurious vacuum energy-momentum of $\varphi$ in flat empty space. Unlike $\left<0\right| \widehat{T}_{\mu\nu}\left|0\right>$, we expect  $T^{\mathrm{Casimir}}_{\mu\nu}$ to be \emph{observable}, giving rise to measurable forces on physical objects, and acting as a source of gravity in the Einstein field equations.

There still remains the technical issue of regularising the two infinite expectation values on the right-hand side of (\ref{CasTdef}), and the question of whether their difference remains finite once the regulator is sent to infinity; for the sake of expediency, however, let us postpone this discussion for now, and take this formal definition of $T^{\mathrm{Casimir}}_{\mu\nu}$ as sufficient for the time being.

As is typical, we will choose the quantum field $\varphi$ to be a free real scalar field. Although correct physical predictions may ultimately require the full complement of standard model fields, it clearly serves no purpose to burden the present abstract investigation with such a detailed and realistic model. Rather, it is hoped that the results of the scalar case will accurately portray the \emph{flavour} of a more complete calculation. In the interest of generality, we will initially proceed without fixing the field's mass; however, as the Casimir effect is exponentially suppressed for systems much larger than a field's Compton wavelength \citep[\S4.2]{Plunien86}, the \emph{massless} case will be our primary interest. The most physically pertinent case will then be the \emph{conformally coupled} massless scalar field, due to the strong analogy with electromagnetism; however, again for the sake of generality, we will leave the curvature coupling parameter arbitrary for now. We begin by summarising the basic ingredients of field theory that we require.

\subsection{Basic Formalism}
The action for the free real scalar field $\varphi$ is
\begin{align}\label{Action}
S_\varphi = \frac{1}{2}\int\! \ud x^4 \sqrt{-g}\left(\left(\nabla \varphi\right)^2 +(m^2+ \xi R) \varphi^2\right),
\end{align}
where $\xi$ is the curvature coupling parameter. For the conformally coupled scalar field, $\xi=1/6$. This action gives rise to the classical field equation
\begin{align}\label{FEq}
0=\frac{-1}{\sqrt{-g}}\frac{\delta S_\varphi}{\delta \varphi}=\left(\nabla^2-m^2-\xi R\right)\varphi,
\end{align}
and the classical energy-momentum tensor
\begin{align}\nonumber
T_{\mu\nu}&\equiv \frac{2}{\sqrt{-g}}\frac{\delta S_\varphi}{\delta g^{\mu\nu}}\\\nonumber
&= \nabla_\mu \varphi\nabla_\nu \varphi + \xi \left(R_{\mu\nu}\varphi^2 -\nabla_\mu \nabla_\nu \left(\varphi^2\right) \right) \\\label{Tclassical}
&\quad\, - g_{\mu\nu}\frac{1-4\xi}{2}\left(\left(\nabla\varphi \right)^2+ (m^2 + \xi R) \varphi^2\right),
\end{align}
where we have used (\ref{FEq}) to simplify the last line. Note that it is only for the conformally coupled field that $T_{\mu\nu}$ agrees with the ``new improved'' energy-momentum tensor which behaves well in the renormalised quantum theory \cite{Callan70}. It will also be convenient to define a symmetric bilinear form $T_{\mu\nu}\left\{\cdot\, ,\cdot \right\}$ based on the classical energy-momentum tensor:
\begin{align}\nonumber
&T_{\mu\nu}\left\{\varphi_1,\varphi_2\right\}\\\nonumber
& \equiv \nabla_{(\mu|} \varphi_{1}\nabla_{|\nu)} \varphi_{2} + \xi \left(R_{\mu\nu}\varphi_1\varphi_2 -\nabla_\mu \nabla_\nu(\varphi_1\varphi_2)  \right) \\\label{Tbi}
&\quad\,- g_{\mu\nu}\frac{1-4\xi}{2}\left(\nabla_\alpha\varphi_1 \nabla^\alpha \varphi_2 +(m^2+ \xi R) \varphi_1 \varphi_2\right).
\end{align}

To quantise $\varphi$, let us set $\hbar=1$ and specialise to ultrastatic spacetimes:
\begin{align}
\ud s^2 \equiv g_{\mu\nu}\ud x^\mu x^\nu = -\ud t^2 + h_{ij}(\vec{x}) \,\ud x^i\ud x^j.
\end{align}
Under canonical quantisation, $\varphi$ is replaced by the operator
\begin{align}\label{modesum}
\widehat{\varphi} = \sum_n \left(\varphi^-_{n}a^-_{n} +\varphi^+_{n}a^{+}_{n} \right),
\end{align}
where the $a^{+}_{n}=(a^{-}_n)^\dagger$ are creation/annihilation operators:
\begin{align}\label{aa}
\left[a_n^+,a_m^+\right]&=\left[a_n^-,a_m^-\right]=0,& \left[a_{n}^-, a^+_{m} \right]&= \delta_{nm}.
\end{align}
In this generic treatment, the mode index $n$ is discrete; for each continuous index $k$ taking values in $\mathbb{R}$, the sum in (\ref{modesum}) should be augmented by $\int\!\ud k/2\pi$, and the Kronecker delta in (\ref{aa}) should be multiplied by $2\pi \delta(k-k^\prime)$. 

The modes $\{\varphi^{\pm}_{n}\}$ are an orthogonal basis of solutions to the field equation (\ref{FEq}); they are required to have definite energy,
\begin{align}
\partial_t \varphi_n^{\pm}&=\pm i \omega_n \varphi_n^{\pm},& \omega_n&\ge0,
\end{align}
and also obey 
\begin{align}\label{realc}
\varphi^{+}_{n}= (\varphi^{-}_{n})^{*},
\end{align}
so that $\hat{\varphi}$ is hermitian (corresponding to $\varphi\in \mathbb{R}$). The modes are normalised such that
\begin{align}\label{cancom}
\left[\widehat{\varphi} (t,\vec{x}), \widehat{\varPi} (t, \vec{x}^\prime)\right]= i h^{-1/2}\delta(\vec{x}-\vec{x}^\prime),
\end{align}
where $\widehat{\varPi}\equiv \partial_t \widehat{\varphi}$ is the conjugate momentum of $\widehat{\varphi}$, and $h\equiv \det (h_{ij})$. Substituting (\ref{modesum}) and using equations (\ref{aa})--(\ref{realc}), this condition becomes
\begin{align}\label{norm}
\sum_n 2 \omega_n \Re\!\left\{\varphi^{+}_n(t,\vec{x})\varphi^{-}_n(t,\vec{x}^\prime)\right\} = h^{-1/2}\delta(\vec{x}-\vec{x}^\prime).
\end{align}
The Fock space is constructed in the usual fashion, with the vacuum state $\left|0\right\rangle$ defined by $a^{-}_{n}\left|0\right\rangle=0$ for all $n$. Consequently, the vacuum energy-momentum tensor is
\begin{align}\nonumber
\left\langle0\right| \widehat{T}_{\mu\nu}\left|0\right\rangle&= \left\langle0\right| T_{\mu\nu}\left\{\widehat{\varphi},\widehat{\varphi} \right\}\left|0\right\rangle\\\nonumber
&= \sum_{n,m} T_{\mu\nu}\left\{ \varphi^-_{n},\varphi^+_{m} \right\}\left\langle0\right|a_{n}^{-}a_{m}^+ \left|0\right\rangle\\\label{Tmodesfless}
&= \sum_{n} T_{\mu\nu}\left\{ \varphi^{-}_{n},\varphi^{+}_{n}\right\}.
\end{align}
Typically this sum will diverge, so some form of regularisation is required to render it meaningful. The simplest approach is to introduce an energy cut-off:
\begin{align}\label{Tmodes}
\left\langle0\right| \widehat{T}_{\mu\nu}\left|0\right\rangle&= \sum_{n} T_{\mu\nu}\left\{ \varphi^{-}_{n},\varphi^{+}_{n}\right\} f(\omega_n/\Omega),
\end{align}
where $f(x)$ is a monotonically decreasing function of $x$, such that $f(0)=1$, which vanishes fast enough as $x\to \infty$ to render the sum finite. Based as it is on the \emph{energy} of the modes, this scheme can be expected to break Lorentz invariance; as such it will be a temporary measure, necessary at this stage to prevent us from deriving nonsense from infinite expressions. In section \ref{Reg}, we will replace it with a Lorentz invariant regularisation scheme and send $\Omega \to \infty$. 

\subsection{Modes in the Infinite Throat}
Fixing the metric to be that of the infinite throat (\ref{throat}), the most convenient set of orthogonal field modes becomes
\begin{align}\label{modes}
\varphi^+_{klm}=(\varphi^-_{klm})^*=\frac{1}{a\sqrt{2\omega}}e^{i (wt-kz)}Y_{lm}(\theta,\phi),
\end{align} 
where $Y_{lm}$ are spherical harmonics, $k\in\mathbb{R}$, $l\in \mathbb{N}$, $m\in \{-l,-l+1,\ldots, l\}$, and 
\begin{align}\label{omega}
\omega \equiv \sqrt{k^2 +m^2+ \frac{l(l+1)+2\xi}{a^2}}
\end{align}
ensures that the field equation (\ref{FEq}) is satisfied. Note that if $2\xi <- m^2 a^2$, then we must disregard modes for which $a^2(k^2+m^2) +l(l+1) +2\xi< 0$, as $\omega$ is imaginary in this case. For now, let us proceed under the assumption that $2\xi \ge- m^2 a^2$; it will be trivial to deal with $2\xi <- m^2 a^2$ in section \ref{Result}, by taking the real part of our results.

To ensure the modes are correctly normalised, we must check that they are in agreement with (\ref{norm}):
\begin{align}\nonumber
\int\frac{\ud k}{2\pi}\sum_{l,m} &2 \omega \Re\left\{\varphi^{+}_{klm}(t,\vec{x})\varphi^{-}_{klm}(t,\vec{x}^\prime)\right\}\\\nonumber
&= \frac{1}{a^{2}}\int\frac{\ud k}{2\pi}\sum_{l,m} e^{ik(z-z^\prime)} Y_{lm}(\theta, \phi)Y^{*}{\!\!\!}_{lm}(\theta^\prime,\phi^\prime) \\\nonumber
&= \frac{1}{a^{2}\sin\theta}\delta(z-z^\prime)\delta(\theta-\theta^\prime)\delta(\phi-\phi^\prime)\\
&=h^{-1/2}\delta(\vec{x}-\vec{x}^\prime). 
\end{align}
Thus the canonical commutation relation (\ref{cancom}) is obeyed.

\subsection{Vacuum Energy-Momentum}
To calculate the vacuum energy-momentum tensor, we substitute the modes (\ref{modes}) into equation (\ref{Tmodes}):
\begin{align}\nonumber
&\left\langle0\right| \widehat{T}_{\mu\nu}\left|0\right\rangle=\\\nonumber
& \int\!\frac{\ud k}{2\pi}\sum_{l, m} \frac{f(\frac{\omega}{\Omega})}{2 \omega a^2}\bigg[ \partial_{(\mu|}\!\left(e^{i (wt-kz)}Y_{lm}\right)\partial_{|\nu)}\!\left(e^{-i (wt-kz)}Y^*{\!\!\!}_{lm}\right) \\\nonumber
&\qquad{}+ \xi \left(R_{\mu\nu}|Y_{lm}|^2 - \nabla_\mu\nabla_\nu\left(|Y_{lm}|^2\right)\right)
\\\nonumber
& \qquad{} - g_{\mu\nu}\frac{1-4\xi}{2}\Big(\!\left(-\omega^2 +k^2 +m^2+ \xi R\right)|Y_{lm}|^2\\\label{vacT1}
&\qquad{} + \partial_\alpha Y_{lm} \partial^\alpha Y^*{\!\!\!}_{lm}\Big)\bigg].
\end{align}
We can perform the sum over $m$ by use of the identities
\begin{align}\bs
\sum_{m=-l}^{l} |Y_{lm}|^2&=\frac{2l+1}{4\pi},\\
\sum_{m=-l}^{l} \partial_\mu Y_{lm}\partial_\nu Y^*{\!\!\!}_{lm} &=\frac{2l+1}{4\pi}\cdot\frac{l(l+1)}{2a^2} \Theta_{\mu\nu},\es
\end{align}
where we have introduced the tensor
\begin{align}
\Theta_{\muh\nuh}\equiv  \mathrm{diag}(0,0,1,1),
\end{align}
to represent the angular part of the metric. The result is
\begin{align}\nonumber
&\left\langle0\right| \widehat{T}_{\mu\nu}\left|0\right\rangle= \\\nonumber
&\int_{-\infty}^\infty\!\frac{\ud k}{16\pi^2 a^2}\sum_{l=0}^\infty \frac{(2l+1) f(\frac{\omega}{\Omega})}{\omega}\bigg[ \omega^2 (\partial_\mu t) (\partial_\nu t) \\\nonumber
&{}+ k^2 (\partial_\mu z) (\partial_\nu z) + \Theta_{\mu\nu}\frac{l(l+1) + 2\xi}{2a^2}\\
&{} -  g_{\mu\nu}\frac{1-4\xi}{2}\left(-\omega^2  +k^2+m^2 + \frac{l(l+1) +2\xi}{a^2} \right) \bigg].
\end{align}
Applying (\ref{omega}) this becomes
\begin{align}\nonumber
\left\langle0\right|\widehat{T}_{\mu\nu}\left|0\right\rangle&= \int^\infty_{-\infty}\!\frac{\ud k}{16\pi^2 a^2}\sum_{l=0}^\infty \frac{(2l+1)f(\frac{\omega}{\Omega})}{\omega}\bigg[ \omega^2 (\partial_\mu t) (\partial_\nu t) \\\nonumber
&\quad\,{}+ k^2 (\partial_\mu z) (\partial_\nu z) + \Theta_{\mu\nu}\frac{\omega^2 - k^2 -m^2}{2}\bigg]\\\nonumber
&= \int^\infty_{-\infty}\!\frac{\ud k}{32\pi^2 a^2}\sum_{l=0}^\infty \frac{(2l+1)f(\frac{\omega}{\Omega})}{\omega}\bigg[ \omega^2A_{\mu\nu}\\
&\quad\,{} + k^2 B_{\mu\nu}-m^2 \Theta_{\mu\nu} \bigg],
\end{align}
in which we have introduced the tensors
\begin{align}\bs\label{ABdef}
A_{\muh\nuh} &\equiv \mathrm{diag}\left(2,0,1,1\right),\\ B_{\muh\nuh} &\equiv \mathrm{diag}\left(0,2,-1,-1\right).\es
\end{align}
Lastly, we define the dimensionless quantities
\begin{align}\label{dimless}
u&\equiv ka,&  v&\equiv \omega a,& \lambda&\equiv \Omega a,& \mu &\equiv ma,
\end{align}
and use them to write
\begin{align}\nonumber
&\left\langle0\right| \widehat{T}_{\mu\nu}\left|0\right\rangle=\\
&\int^{\infty}_{0}\!\frac{\ud u}{8\pi^2 a^4}\sum_{l=0}^\infty \frac{(l+\half)f(\frac{v}{\lambda})}{v}\left[ v^2 A_{\mu\nu}+ u^2 B_{\mu\nu}- \mu^2 \Theta_{\mu\nu} \right],\label{throatT}
\end{align}
wherein
\begin{align}\label{alphadef}
v&\equiv\sqrt{u^2 +(l+1/2)^2 + \alpha},&\alpha &\equiv\mu^2+ 2\xi - 1/4 .
\end{align}

\subsection{Minkowski Vacuum Energy-Momentum}
To complete the calculation of $T^\text{Casimir}_{\mu\nu}$, we also require the vacuum energy-momentum of $\varphi$ in Minkowski spacetime (\ref{mink}), evaluated according to the same regularisation scheme. The Minkowski modes are of course
\begin{align}
\varphi^+_{\vec{k}}&=\big(\varphi^-_{\vec{k}}\big)^*= \frac{1}{\sqrt{2\ob}}e^{i(\ob t-\vec{k}\cdot\vec{x})}, &\ob &\equiv\sqrt{ |\vec{k}|^2 +m^2},
\end{align}
and lead to a regularised vacuum energy
\begin{align}\nonumber
\left\langle0_M\right|& \widehat{T}_{\muh\nuh}\left|0_M\right\rangle \\\label{MinkVac0}
&= \int\frac{\ud^3 \vec{k}}{(2\pi)^3}\frac{f(\frac{\ob}{\Omega})}{2\ob}\mathrm{diag}(\ob^2,(k_1)^2,(k_2)^2,(k_3)^2 ).
\end{align}
To rewrite this integral in a way the resembles the throat result (\ref{throatT}) let us parameterise $\vec{k}= (k, q \cos \vartheta, q\sin \vartheta)$ and perform the integral over $\vartheta$:
\begin{align}\nonumber
&\left\langle0_M\right| \widehat{T}_{\muh\nuh}\left|0_M\right\rangle\\\nonumber
& =\int^\infty_{-\infty}\!\frac{\ud k}{(2\pi)^2}\int^\infty_0\!\!\!\ud q q \frac{f(\frac{\ob}{\Omega})}{2\ob}\mathrm{diag}(\ob^2,k^2,q^2/2,q^2/2 )\\\label{MVac0}
& =\int^\infty_{-\infty}\!\frac{\ud k}{16\pi^2}\int^\infty_0\!\!\!\ud q q \frac{f(\frac{\ob}{\Omega})}{\ob}\left[\ob^2 A_{\muh\nuh}+ k^2 B_{\muh\nuh} - m^2 \Theta_{\muh\nuh}\right],
\end{align}
where $q^2= \ob^2-k^2 -m^2$ was used in the last line. Writing $q= l/a$ (with $l$ a continuous variable) we express everything in terms of the dimensionless variables (\ref{dimless}):
\begin{align}\nonumber
&\left\langle0_M\right| \widehat{T}_{\mu\nu}\left|0_M\right\rangle\\\label{MinkT}
&=\int^\infty_{0}\frac{\ud u}{8\pi^2 a^4}\int^\infty_0 \ud l  \frac{lf(\frac{\vb}{\lambda})}{\vb}\left[ \vb^2 A_{\mu\nu}+ u^2 B_{\mu\nu} -\mu^2 \Theta_{\mu\nu} \right],
\end{align}
wherein
\begin{align}\label{alphabardef}
\vb&\equiv \sqrt{u^2 + l^2+ \alphabar}, &\alphabar&\equiv\mu^2.
\end{align}

\subsection{Casimir Energy-Momentum}\label{Result}
Finally, we subtract the Minkowski energy-momentum (\ref{MinkT}) from the throat energy-momentum (\ref{throatT}) to arrive at the Casimir energy-momentum tensor:
\begin{align}
T^{\mathrm{Casimir}}_{\mu\nu}=\frac{1}{8\pi^2 a^4} \left(I A_{\mu\nu} +  J B_{\mu\nu} -\alphabar K \Theta_{\mu\nu}\right),
\end{align}
where 
\begin{align}\nonumber
I& \equiv \int^\infty_0\! \ud u \left[ \sum_{l=0}^\infty(l+\half) v f(\tfrac{v}{\lambda}) - \int^\infty_0 \!\ud l \, l \vb f(\tfrac{\vb}{\lambda})\right],\\\nonumber
J &\equiv \int^\infty_0\! \ud u \Bigg[ \sum_{l=0}^\infty \frac{(l+\half)u^2f(\frac{v}{\lambda})}{v}- \int^\infty_0 \!\ud l \, \frac{l u^2 f(\frac{\vb}{\lambda})}{\vb}\Bigg],\\\nonumber
K&\equiv \int^\infty_0\! \ud u \Bigg[ \sum_{l=0}^\infty \frac{(l+\half)f(\frac{v}{\lambda})}{v}- \int^\infty_0 \!\ud l \, \frac{l f(\frac{\vb}{\lambda})}{\vb}\Bigg],\\\nonumber
v&\equiv\sqrt{u^2 +(l+1/2)^2 + \alpha},\quad \alpha \equiv\mu^2+ 2\xi - 1/4,\\
\vb&\equiv \sqrt{u^2 + l^2+ \alphabar},\qquad\qquad\ \,  \alphabar\equiv\mu^2.
\end{align}
Recall that this result is only valid for $2\xi \ge - m^2 a^2$ (equivalently, $\alpha \ge - 1/4 $) and that if $2\xi < - m^2 a^2$ we must be careful to remove any modes for which $\omega$ is imaginary. Fortunately, these modes produce a purely imaginary contribution to $I$, $J$ and $K$, so they are easily removed simply by taking the real part of the above expression. Hence,
\begin{align}
T^{\mathrm{Casimir}}_{\mu\nu}&=\frac{1}{8\pi^2 a^4}\Re \left\{I A_{\mu\nu} +  J B_{\mu\nu}-\alphabar K \Theta_{\mu\nu}\right\}
\end{align}
is now valid for all $\alpha\in \mathbb{R}$.

In appendix \ref{Calc}, we use the Abel-Plana formula to evaluate $\Re \left\{I \right\}$, $\Re \left\{J\right\}$ and $\Re \left\{K\right\}$ when $f$ enacts a \emph{sharp} cut-off at energy $\Omega=\lambda/ a$; the results can be found in equations (\ref{Iresult}-\ref{Kresult}). Consequently, the Casimir energy-momentum tensor (expressed in the orthonormal basis) is given by
\begin{widetext}
\begin{align}\nonumber
T^{\mathrm{Casimir}}_{\muh\nuh}&=\frac{1}{8\pi^2 a^4}\big[\Re \{I+J\}\mathrm{diag}(1,1,0,0) +  \Re \{I-J\}\mathrm{diag}(1,-1,1,1) - \alphabar \Re \{K\}\mathrm{diag}(0,0,1,1) \big]\\\nonumber
&= \frac{1}{192\pi^2 a^4}\Bigg[\left(\left(1+ 12(\alphabar-\alpha)\right)\lambda^2  -\frac{1}{2}\left(\alpha +6(\alphabar^2-\alpha^2) -\frac{7}{40}\right)\right)\mathrm{diag}(1,1,0,0)\\\nonumber
&\quad\,{}+\left(\left(\alpha +6(\alphabar^2-\alpha^2) -\frac{7}{40}\right)\ln(2\lambda) + 3(\alpha^2 \ln|\alpha|-\alphabar^2 \ln|\alphabar|) - \frac{3}{2}(\alpha^2-\alphabar^2) +48X(\alpha)\right)\mathrm{diag}(1,-1,1,1)\\\label{BigT}
&\quad\,{}-\alphabar\Big((1+12(\alphabar-\alpha)) \ln(2\lambda)+6\left(\alpha \ln |\alpha| - \alphabar\ln |\alphabar|\right) +6(\alphabar - \alpha) - 48Y(\alpha) \Big)\mathrm{diag}(0,0,1,1)\Big] + O(\lambda^{-2}),
\end{align}
\end{widetext}
as $\lambda\to \infty$, where the functions $X(\alpha)$ and $Y(\alpha)$ are defined in equations (\ref{Xdef}) and (\ref{Ydef}).

At this stage, the key fact to recognise is that, unlike in Casimir's original calculation \cite{Casimir48}, we cannot send $\lambda\to\infty$ and recover a finite result: although the subtraction of the Minkowski vacuum has removed a singularity $O(\lambda^4)$ from the expansion, there remains singularities $O(\lambda^2)$ and $O(\ln \lambda)$.\footnote{We can nullify the $\lambda^2$ divergence by setting $\alpha-\alphabar=1/12$, which corresponds to conformal coupling $\xi=1/6$; however, the coefficient of the remaining $\ln(\lambda)$ term is then $\alpha +6(\alphabar^2-\alpha^2) -7/40= -2/15$, independent of the value of $\mu^2$, and so cannot also be set to zero.} With finite $\lambda$, Lorentz invariance remains broken, and this is manifest in two ways. Firstly, $T^{\mathrm{Casimir}}_{\mu\nu}$ is dependent on the cut-off energy $\Omega= \lambda/a$, which is obviously a frame-dependent quantity. Secondly, the piece of $T^{\mathrm{Casimir}}_{\mu\nu}$  proportional to $\mathrm{diag}(1,1,0,0)$ is not invariant under boosts in the $z$-direction, and so does not respect the Lorentz symmetry of the throat metric (\ref{throat}). The purpose of the following section is to remedy these failings with the introduction of a Lorentz-invariant regularisation scheme.

\section{Restoring Lorentz Invariance}\label{Reg}
The simplest way to restore Lorentz-invariance to $T^{\mathrm{Casimir}}_{\muh\nuh}$ is to introduce a Pauli-Villars regulator \cite{Pauli49}. The regulator is a fictitious scalar field $\varphi_*$ (with a very large mass $m_*$) the energy-momentum of which we subtract from that of $\varphi$:
\begin{align}\label{PVdef}
T^{\mathrm{PV}}_{\mu\nu}\equiv T^{\mathrm{Casimir}}_{\mu\nu}[\varphi] -T^{\mathrm{Casimir}}_{\mu\nu}[\varphi_*].
\end{align}
Following this scheme, the low-energy modes of $\varphi$ contribute to $T^{\mathrm{PV}}_{\mu\nu}$ as usual, with negligible subtraction from $\varphi_*$; for modes with energies far above $m_*$, however, the contributions from the two fields almost exactly cancel. Consequently, high-energy modes are suppressed in a smooth and Lorentz-invariant fashion. Once this regulator has been added, it will hopefully be possible to send our original cut-off $\lambda\to \infty$ , with $T^{\mathrm{PV}}_{\mu\nu}$ remaining finite, and $m_*$ retained as a  Lorentz-invariant regularisation scale.

Although Pauli-Villars regularisation is rarely used for Casimir energy-momentum calculations, it has a number of advantages over the alternative schemes (dimensional regularisation \cite{Bollini72,'tHooft72}, point splitting \cite{Adler76}, and zeta-function regularisation \cite{Dowker76,Hawking77}) at least for the case at hand. Firstly, this approach follows very easily from the energy cut-off result (\ref{BigT}), requiring only elementary algebra, with no need for additional mathematical tools or formalism. Secondly, the Pauli-Villars approach has no additional ambiguities or freedoms, beyond the energy-scale $m_*$ that is present in all methods.\footnote{For instance, there is the question of how, in detail, one should  perform dimensional regularisation: the wormhole throat (\ref{throat}) has a (wick-rotated) geometry $\mathbb{R}^2\times S^2$ which generalises to $\mathbb{R}^{d_1}\times S^{d_2}$ with two degrees of freedom. Similarly, point splitting requires a choice of \emph{splitting direction}, and the subtleties in application of the zeta-function give rise to other ambiguities \citep[p167]{BirrellDavies}. Even if these ambiguities can be fixed \emph{post hoc} (e.g.\ by insisting on some property of $T_{\mu\nu}$, or by averaging over splitting direction \cite{Adler76}) or can be absorbed in the process of renormalisation, it will clearly be advantageous to avoid these additional complications.} Lastly, Pauli-Villars has an attractive ``toy model'' physical interpretation: one can think of the regulator field as representing the appearance of new particle species (at an energy scale $m_*$) which suppress the energy-momentum of high-energy modes of $\varphi$. This behaviour would be expected from spontaneously broken supersymmetry: as energies exceed the symmetry-breaking scale, superpartner fields would appear that exactly cancel the energy-momentum contribution of the fields present at low energy. Of course, this is only a toy model, and one may prefer to default to the minimal interpretation, wherein regularisation is a purely mathematical device, devoid of physical meaning. As we will see in section \ref{Renorm}, this more rigorous approach (in which $m_*$ is eventually sent to infinity and divergences are absorbed via renormalisation) gives essentially the same results as the toy model.

\subsection{Pauli-Villars Regularisation}
Employing the Pauli-Villars regularisation scheme is simply a matter of inserting (\ref{BigT}) into (\ref{PVdef}). Although it is certainly possible to proceed without fixing the mass of $\varphi$, it will streamline our analysis to now focus specifically on the massless case. Recall that this was our original intention: the Casimir effect of the massive case is expected to diminish as $e^{-2ma}$ for $ma\gg 1$ \citep[\S4.2]{Plunien86}, and so would be unable to support a macroscopic wormhole. Of course, we will still require the formula (\ref{BigT}) for $m\ne0$ in order to calculate the contribution from the regulator field.

Referring to definitions (\ref{dimless}), (\ref{alphadef}) and (\ref{alphabardef}), we see that the massless field $\varphi$ requires $\alpha= 2\xi-1/4$ and $\alphabar =0$, and the regulator $\varphi_*$ requires $\alpha= 2\xi-1/4 + (m_*a)^{2}$ and $\alphabar =(m_*a)^{2}$. Hence, the Pauli-Villars regularised Casimir energy-momentum (\ref{PVdef}) can be written as
\begin{align}\label{PVdef2}
\!\!\! T^{\mathrm{PV}}_{\mu\nu}&= \left.T^{\mathrm{Casimir}}_{\mu\nu}\right|_{\alpha=\zeta,\,\alphabar=0}-\left.T^{\mathrm{Casimir}}_{\mu\nu}\right|_{\alpha=\zeta+ \mu_*^2,\,\alphabar=\mu_*^{2}},
\end{align}
where
\begin{align}\label{shorthand}
\zeta& \equiv2 \xi -1/4,&
\mu_*& \equiv m_*a,
\end{align}
have been introduced as a convenient shorthand. We can now substitute (\ref{BigT}) into (\ref{PVdef2}) and arrive at
\begin{widetext}
\begin{align}\nonumber
T^{\mathrm{PV}}_{\muh\nuh}&= \frac{1}{192\pi^2 a^4}\Bigg[\frac{\mu_*^2}{2}\left(1-12\zeta \right)\mathrm{diag}(1,1,0,0) + \Big(\mu_*^2(12\zeta -1)\ln(2\lambda)+ 3 \zeta^2 (\ln |\zeta| - 1/2) - 3(\mu_*^2+ \zeta)^2(\ln |\mu_*^2+ \zeta | - 1/2)
\\\nonumber
&\quad\,{}  +  3\mu_*^4(\ln |\mu_*^2| - 1/2)  + 48X(\zeta) - 48X(\mu_*^2+ \zeta )\Big)\mathrm{diag}(1,-1,1,1) + \mu_*^2 \Big((1-12\zeta)\ln(2\lambda) + 6(\mu_*^2+ \zeta)\ln |\mu_*^2+\zeta|
\\\label{PV1}
&\quad\,{} -6 \mu_*^2\ln |\mu_*^2| - 6 \zeta - 48 Y(\mu_*^2+\zeta) \Big)\mathrm{diag}(0,0,1,1)\Bigg] + O(\lambda^{-2}).
\end{align}
\end{widetext}
Notice that, as a consequence of the new regularisation, the $O(\lambda^2)$ divergence has vanished entirely. In general, though, two pathologies still remain: firstly, a divergence $O(\ln(\lambda))$, and secondly, the frame-dependent tensor $\mathrm{diag}(1,1,0,0)$. Fortunately, both of these features can be removed simply by setting $\zeta=1/12$; a glance at (\ref{shorthand}) confirms that this is the same as
\begin{align}\label{conformal}
\xi=1/6,
\end{align}
which of course ensures that $\varphi$ is \emph{conformally coupled}. Recall that this value for $\xi$ was physically well-motivated \emph{a priori}, as the conformal invariance of $\varphi$ makes it most closely analogous to the electromagnetic field, and thus a good model for the only massless field in the standard model. Moreover, (\ref{conformal}) also defines the ``new improved'' energy-momentum tensor \cite{Callan70} of Callan, Coleman and Jakiw, which is the form of energy-momentum tensor most suitable for renormalised quantum theory. Thus, although the other possibilities $\xi\ne1/6$ can presumably be dealt with using a more complicated regularisation scheme, there seems little to be gained in pursuing these results, considering that they were less well-motivated in the first place.

With conformal coupling (\ref{conformal}) fixed, there is nothing to stop us from sending $\lambda \to\infty$ and recovering Lorentz invariance. The final step of the calculation is then to let our Lorentz-invariant regulator $m_*$ become very large; specifically, we insist that the wormhole radius should be much greater than the regulator's Compton wavelength, so $m_* a=\mu_*\gg 1$. We can then use
\begin{align}
\ln |\mu_*^2+ \zeta |&= \ln |\mu_*^2| + \frac{\zeta}{\mu_*^2} - \frac{\zeta^2}{2\mu_*^4} + O(\mu_*^{-6}),
\end{align}
and the following asymptotic expansions,
\begin{align}\nonumber
X(\mu_*^2+ \zeta ) &= \frac{1}{96}\left( \left(\frac{7}{40}- \zeta -\mu_*^2 \right)\ln |\mu_*^2| - \zeta + \frac{7}{40}\right)\\
&\quad\, {} + O(\mu_*^{-2}),\\\nonumber
Y(\mu_*^2+ \zeta ) &=  \frac{1}{96} \left(\ln |\mu_*^2| + \frac{1}{\mu_*^2}\left(\zeta - \frac{7}{40}\right)  \right) + O(\mu_*^{-4}),
\end{align}
which can be derived by the same methods as used in the steps between (\ref{logtaylor}) and (\ref{I1}). Inserting these into (\ref{PV1}) with $\zeta = 1/12$, we finally obtain (after considerable cancelling) the Pauli-Villars regularised Casimir energy-momentum tensor: 
\begin{align}\nonumber
T^{\mathrm{PV}}_{\muh\nuh}&= \frac{1}{2880 \pi^2 a^4}\big[ \mathrm{diag}(-1,1,-1,-1)(\ln|\mu_*^2| + \Delta)\\\label{PVresult}
&\qquad\qquad\qquad\,{}  +  \mathrm{diag}(0,0,1,1)\big] + O(\mu_*^{-2}),
\end{align}
where
\begin{align}\nonumber
\Delta &\equiv \frac{37 + 10\ln(12)}{32} - 360 \int^{\infty}_{0}\!\ud t  \, \frac{t(t^2-\tfrac{1}{12})\ln |t^2 -\tfrac{1}{12}|}{e^{2\pi t} +1}\\
&\cong 2.2325
\end{align}
is a numerical factor.

As far as the toy model is concerned, we can end the calculation here. In (\ref{PVresult}) we have $\mu_*=m_*a$, where $m_*$ is a very large, unknown but finite mass,  quantifying the energy at which new particle species arise and suppress the energy-momentum of $\varphi$. Our present analysis cannot predict $m_*$, but it could be determined experimentally, at least in principle. Fortunately, as $T^{\mathrm{PV}}_{\mu\nu}$ is only logarithmically dependent on $m_*$, this uncertainty will have little impact on our conclusions. In particular, we know that $\mu_*\gg 1$, so $\ln|\mu_*^2|$ is positive, and thus the Casimir energy-density $T^{\mathrm{PV}}_{00}$ is \emph{negative}, just as we had hoped. However, negative energy-density is not sufficient in itself to guarantee the stability of the wormhole; we will examine this subject properly in section \ref{Stab}.

Before this, it will be useful to briefly review how renormalisation allows us to take the regulator $m_*$ to infinity, while retaining a finite energy-momentum tensor. In truth, this more rigorous treatment has little effect on the important features of $T^{\mathrm{PV}}_{\mu\nu}$, so a reader who is happy to accept the toy model picture of Pauli-Villars regularisation may wish to skip to section \ref{Stab} at this point. In section \ref{Therm} we will also discuss the conformal anomaly displayed by (\ref{PVresult}).

\subsection{Renormalisation}\label{Renorm}
Prior to renormalisation, the semi-classical Einstein field equations are
\begin{align}\label{BareEFE}
G_{\mu\nu} + g_{\mu\nu}\Lambda_\mathrm{B} = \kappa_\mathrm{B} \left< T_{\mu\nu}\right>,
\end{align}
with  `bare' cosmological and gravitational constants $\Lambda_\mathrm{B}$ and $\kappa_\mathrm{B}$. Utiyama and DeWitt \cite{Utiyama62} proved that the expectation value of the energy-momentum tensor will generically take the form
\begin{align}\nonumber
\left< T_{\mu\nu}\right> &= c_1 g_{\mu\nu} m_*^4 + c_2G_{\mu\nu}  m_*^2 + c_3  H_{\mu\nu}\ln(m_* b)\\\label{Texpand}
&\quad{} + T^\mathrm{ren}_{\mu\nu},
\end{align}
where $m_*$ is a Lorentz-invariant regulator, $\{c_1, c_2, c_3\}$ are numerical constants, $H_{\mu\nu}$ is a tensor composed of $R^2$ and $\nabla^2 R$ terms,\footnote{In fact, there are two linearly independent terms of this sort, so $c_3 H_{\mu\nu}$ should really be replaced by $c_3 H_{\mu\nu}^{(1)}+ c_4 H_{\mu\nu}^{(2)}$. This complication is irrelevant to the schematic explanation given here, so we will ignore it.} and $T^\mathrm{ren}_{\mu\nu}$ is finite as $m_*\to \infty$.  Notice that it has been necessary to introduce an \emph{arbitrary} length-scale $b$, without which the logarithm would have a dimensionful argument.  Because $\left< T_{\mu\nu}\right>$ does not actually depend on $b$, any change $b\to b^\prime$ must produce a compensating change in the finite part of the energy-momentum tensor: $\Delta T^\mathrm{ren}_{\mu\nu}=  c_3 \ln (b/b^\prime) H_{\mu\nu}$.

Substituting (\ref{Texpand}) into (\ref{BareEFE}), grouping terms and dividing by $(1-c_2 \kappa _\mathrm{B}m_*^2)$, we arrive at the following field equations:
\begin{align}\nonumber
G_{\mu\nu} + g_{\mu\nu}&\frac{\Lambda_\mathrm{B} - c_1 m_*^4\kappa_\mathrm{B}}{1-c_2 \kappa _\mathrm{B}m_*^2}\\\label{ren1}
 &=  \frac{\kappa_\mathrm{B}}{1-c_2 \kappa _\mathrm{B}m_*^2}\Big[ T^\mathrm{ren}_{\mu\nu}  + c_3  H_{\mu\nu}\ln(m_* b)\Big].
\end{align}
Ignoring the logarithmic divergence for the moment, we see that neither the bare constants $\Lambda_\mathrm{B}, \kappa_\mathrm{B}$, nor the $m_*^2, m_*^4$ divergences, can be observed directly: one can only measure the renormalised quantities
\begin{align}\label{renconst}
\Lambda &\equiv \frac{\Lambda_\mathrm{B} - c_1 m_*^4\kappa_\mathrm{B}}{1-c_2 \kappa _\mathrm{B}m_*^2},&  \kappa &\equiv\frac{\kappa_\mathrm{B}}{1-c_2 \kappa _\mathrm{B}m_*^2}.
\end{align}
These quantities have been measured experimentally, and are known to be finite.\footnote{The empirical value of the cosmological constant is so small that it is unlikely to have a significant effect on the wormhole; as such, we set $\Lambda=0$ outside this section of the paper.} Consequently, we can infer the behaviour of $\Lambda_\mathrm{B}$ and $\kappa_\mathrm{B}$ as $m_*\to \infty$.

This deals with the quadratic and quartic divergences: they simply produce  an unobservable shift in the cosmological and gravitational constants.\footnote{As it happens, neither of these divergences are present in (\ref{PVresult}): the quadratic divergence does not appear when the field is conformally coupled, and the quartic divergence was removed by subtracting the Minkowski energy-momentum in equation (\ref{CasTdef}). This latter process is essentially a renormalisation of $\Lambda$.} The interesting physical behaviour is then confined to $T^\mathrm{ren}_{\mu\nu}$ and the logarithmic divergence. To absorb this divergence, one must posit the existence of extra $R^2$ terms in the gravitational action, with an (unobservable) bare coupling parameter $\sigma_\mathrm{B}$. These contributions produce a term $\sigma_\mathrm{B}H_{\mu\nu}/(1-c_2 \kappa _\mathrm{B}m_*^2)$ on the left-hand-side of the field equations (\ref{ren1})  which combines with the logarithmic divergence to give
\begin{align}
G_{\mu\nu} + g_{\mu\nu}\Lambda + \sigma H_{\mu\nu}= \kappa T^\mathrm{ren}_{\mu\nu},
\end{align}
where the renormalised coupling
\begin{align}
\sigma\equiv\frac{ \sigma_\mathrm{B}}{1-c_2 \kappa _\mathrm{B}m_*^2} - c_3 \kappa \ln(m_* b)
\end{align}
can once again be determined by experiment. Astrophysical observations provide a stringent upper bound for $\sigma$, so it goes without saying that its value must be finite; indeed, $\sigma=0$ remains a possibility, in which case Einstein's theory survives in spite of the $R^2$ counter-terms in the action.

Notice that the definition of $\sigma$ depends on the arbitrary length-scale $b$. Changing $b$ allows us to add any finite amount to $\sigma$, with an equal and opposite change in $T^\mathrm{ren}_{\mu\nu}$. A particularly convenient way to remove this ambiguity is to choose $b$ such that $\sigma=0$. This convention promotes $b$ to a physically meaningful quantity (determined by experiment) and ensures the field equations take on their usual Einsteinian form,
\begin{align}\label{EFEs}
G_{\mu\nu} + g_{\mu\nu}\Lambda = \kappa T^\mathrm{ren}_{\mu\nu}.
\end{align}

Let us now apply this process to the Casimir energy-momentum of the long wormhole throat (\ref{PVresult}). We write the divergent logarithm as
\begin{align}\nonumber
\ln|\mu^2_*|=2\ln(m_* a)= 2\ln(m_*b) +2 \ln (a/b),
\end{align}
with the first piece identified as producing the logarithmic divergence in (\ref{Texpand}), and the second piece to be included in $T^\mathrm{ren}_{\mu\nu}$. Following the scheme above, $R^2$ terms are added to the gravitational action, the logarithmic divergence is absorbed into $\sigma H_{\mu\nu}$, and we fix $b$ by insisting that $\sigma=0$. Consequently, we arrive at the renormalised Casimir energy-momentum tensor 
\begin{align}\nonumber
T^\mathrm{ren}_{\muh\nuh}&=\frac{1}{2880 \pi^2 a^4}\big[ \mathrm{diag}(-1,1,-1,-1)2 \ln(a/a_0)\\\label{Tren}
&\qquad\qquad\qquad\,{}  +  \mathrm{diag}(0,0,1,1)\big],
\end{align}
where 
\begin{align}
a_0\equiv b e^{-\Delta/2}
\end{align}
is a fixed length that can only be determined by experiment.

Equation (\ref{Tren}) is then our final result. As previously advertised, the renormalised energy-momentum tensor displays much of the same structure as the Pauli-Villars regularised tensor (\ref{PVresult}), with the unknown length-scale $a_0$ replacing $1/m_*$. The parameter $a_0$ has a straightforward interpretation: it is the throat-radius of a wormhole for which the Casimir energy-density vanishes. Provided the wormhole has a throat-radius \emph{greater} than $a_0$, the Casimir energy-density will be \emph{negative}.

\subsection{Conformal Anomaly and Wormhole Thermodynamics}\label{Therm}
Being largely irrelevant to the stability of the wormhole, we have thus far paid little attention to the $\mathrm{diag}(0,0,1,1)$ part of the Casimir energy-momentum tensor (\ref{Tren}). However, this part of the tensor plays a key role in generating the conformal anomaly, and ensuring the self-consistent thermodynamic behavior of the wormhole. We shall quickly cover these details here, for the sake of completeness, before finally examining the energy conditions violated by $T^\mathrm{ren}_{\mu\nu}$.

The presence of a conformal anomaly is evidenced by the trace of the renormalised energy-momentum tensor, 
\begin{align}\nonumber
T^\mathrm{ren}= \frac{1}{1440\pi^2a^4},
\end{align}
which classically would be expected to vanish for a conformally coupled massless scalar field. This anomaly could have been anticipated from general considerations of quantum fields in curved backgrounds; for example, one might have used equation (6.114) of Birrell \& Davies \cite{BirrellDavies}. Accounting for the difference in metric sign convention, this gives
\begin{align}\nonumber
T^\mathrm{ren}&= \frac{1}{2880\pi^2}\big(R_{\alpha\beta\gamma\delta}R^{\alpha\beta\gamma\delta} - R_{\alpha\beta}R^{\alpha\beta} - \nabla^2 R \big)\\
&= \frac{1}{1440\pi^2a^4},
\end{align}
in agreement with the result above. This trace, which arises from the $\mathrm{diag}(0,0,1,1)$ part of $T^\mathrm{ren}$, is intimately connected to the logarithmic dependence of the traceless part (proportional to $\mathrm{diag}(-1,1,-1,-1)$) as we will see by examining the thermodynamical behaviour of the wormhole.

Consider a section of throat (\ref{throat}) of length $l$. From (\ref{Tren}) we see that Casimir energy contained within is 
\begin{align}\label{E}
E = 4 \pi a^2 l \rho = -\frac{ l }{360\pi a^2}\ln(a/a_0).
\end{align}
Thus, if the throat radius undergoes a change $\ud a$ (with $a_0$ held constant) the Casimir energy is altered by
\begin{align}
\ud E = -\frac{l}{360\pi a^3} (-2\ln(a/a_0) + 1)\ud a,
\end{align}
where the $+1$ arises from differentiating the logarithm. The energy-momentum tensor (\ref{Tren}) also reveals the pressure acting in the angular directions:
\begin{align}\label{P}
P= \frac{1}{2880 \pi^2 a^4} (-2 \ln(a/a_0) +1),
\end{align}
where the $+1$ arises from the $\mathrm{diag}(0,0,1,1)$ part of $T^\mathrm{ren}_{\mu\nu}$. Consequently, under a change in radius, the work done by the throat (on the field $\varphi$) is
\begin{align}\nonumber
P\ud V &=\frac{1}{2880 \pi^2 a^4} (-2 \ln(a/a_0) +1)\ud (4\pi a^2 l)\\
&= \frac{l}{360\pi a^3} (- 2\ln(a/a_0) + 1)\ud a.
\end{align}
Hence the logarithm and the $\mathrm{diag}(0,0,1,1)$ tensor conspire to ensure that the first law of thermodynamics
\begin{align}
\ud E = - P\ud V
\end{align}
is obeyed when the throat radius expands or contracts.

\section{Wormhole Stability}\label{Stab}
When the throat radius is sufficiently large ($a > a_0$) the renormalised Casimir energy-density $T^\mathrm{ren}_{00}$ is negative (in violation of the weak and dominant energy conditions) and has a magnitude of order $a^{-4}$. Hence the required exotic energy-density $-2/L a\kappa$ (obtained from the second term on the right of (\ref{2parT}) in the limit that the throat length $L$ becomes extremely large) can be supplied by the Casimir energy-momentum tensor (\ref{Tren}) if
\begin{align}
\frac{-2}{L a\kappa} = \frac{-\ln(a/a_0)}{1440 \pi^2 a^4},
\end{align}
or equivalently,
\begin{align}\label{a^2}
a^2 = (l_p)^2 (L/a) \frac{\ln (a/a_0)}{360 \pi},
\end{align}
where $l_p$ is the Planck length: $\kappa = 8\pi (l_p)^2$. Thus, as hypothesised in the introduction, the Casimir effect of a very long wormhole ($L\gg a$) is capable of generating exotic matter in quantities that could, in principle, allow for the stability of a wormhole with a \emph{macroscopic} throat radius $a\gg l_p$.

However, thus far we have satisfied only the first component of the Einstein field equation (ignoring nonexotic matter). Unfortunately, the \emph{positive} Casimir pressure $T^\mathrm{ren}_{11}= -T^\mathrm{ren}_{00} $ will prove problematic when extending this procedure to the other components. To understand this issue, let us relax our focus on the single component $T^\mathrm{ren}_{00}$, and consider the inequalities obeyed by the tensor $T^\mathrm{ren}_{\mu\nu}$ as a whole.

\subsection{Energy Conditions}\label{ECs}
As previously mentioned, the weak and dominant energy conditions are clearly violated if $a>a_0$;  indeed, we show in Appendix \ref{EnAp} that if this inequality is strengthened very slightly, so that
\begin{align}\label{a>e}
a > a_0 e^{1/4},
\end{align}
then $T^\mathrm{ren}_{\mu\nu}$ will violate all four energy conditions: weak, strong, dominant and null. Furthermore, the stipulation (\ref{a>e}) ensures that, for all null or timelike vectors $v^\mu$,
\begin{align}\label{exotic}
v^\mu T^\mathrm{ren}_{\mu\nu} v^\nu&\le 0, &( T^\mathrm{ren}_{\mu\nu} v^\nu)^2 \le 0,
\end{align}
with equality if and only if $v^\mu \propto (1,\pm1,0,0)$. The first inequality reveals that all four-velocities $v^\mu$ define \emph{negative} energy-densities, with the sole exception being null vectors running directly parallel to the throat, for which the energy-density is zero. The second inequality reassures us that the energy-flux is always \emph{causal}: $T^\mathrm{ren}_{\mu\nu} v^\nu$ is never spacelike. Thus $T^\mathrm{ren}_{\mu\nu}$ is impressively exotic (defining negative energy-densities in almost all directions) but does not violate the more fundamental expectation that energy (whether positive or negative) should never flow faster than light.

With regards to the stability of the wormhole, the problem arises from those particular null directions $v^\mu \propto (1,\pm1,0,0)$ which define vanishing energy-density. We would like to be able to solve the  Einstein equations with the addition of some ordinary matter $T^\mathrm{ord}_{\mu\nu}$:
\begin{align}\label{G=T+T}
G_{\mu\nu}= \kappa (T^\mathrm{ren}_{\mu\nu}+T^\mathrm{ord}_{\mu\nu}),
\end{align}
where $G_{\mu\nu}$ is given by (\ref{2parT}) in the large $L$ limit. Let us contract this equation with $v^\mu v^\nu$, where $v^\mu$ is timelike or null: 
\begin{align}
v^\mu G_{\mu\nu}v^\nu= \kappa v^\mu(T^\mathrm{ren}_{\mu\nu}+T^\mathrm{ord}_{\mu\nu})v^\nu.
\end{align}
For all timelike $v^\mu$, and almost all null $v^\mu$, we have $v^\mu T^\mathrm{ren}_{\mu\nu} v^\nu < 0$, so we should be able to accommodate negative values for $v^\mu G_{\mu\nu}v^\mu$. However, for $v^\mu \propto (1,\pm1,0,0)$, we have $v^\mu T^\mathrm{ren}_{\mu\nu}v^\mu=0$ and $v^\mu G_{\mu\nu}v^\mu<0$, leaving us with 
\begin{align}
v^\mu T^\mathrm{ord}_{\mu\nu}v^\nu <0,
\end{align}
which requires the ordinary matter to violate the null energy condition -- clearly a contradiction. Thus it is impossible to solve the Einstein equations for the static wormhole (\ref{2par}) in the large $L$ limit, using only its Casimir energy-momentum (\ref{Tren}) and ordinary matter. Unfortunately, the exotic energy-momentum generated is not quite of the right form to stabilise the wormhole. 

The root cause of this obstacle is the symmetry of the throat metric (\ref{throat}) under Lorentz boosts in the $z$-direction. Consistency with this symmetry (and the spherical symmetry of the cross-sections) guarantees that $T^{\mathrm{ren}}_{\muh\nuh}= \mathrm{diag}(\rho,-\rho,p,p)$, for some $\rho$ and $p$, with $v^\mu \propto (1,\pm1,0,0) \Rightarrow v^\mu T^\mathrm{ren}_{\mu\nu}v^\mu=0$ an inevitable consequence. Considering that all other four-velocities define negative energy-density, it is conceivable that some symmetry-breaking deformation of the spacetime may alleviate this problem. In particular, noting that null rays with nonzero angular momentum \emph{do} see negative energy-density, one might hope that introducing a ``twist'' to the throat would mix the angular and longitudinal behaviour in a profitable way. Alternatively, one could examine wormholes with very \emph{short} throats: their extreme aspect ratios leave them open to the method of attack described in the introduction, but unlike the $L\to\infty$ limit considered here, the $L\to 0$ limit is not invariant under longitudinal boosts. We shall leave these possibilities for separate investigations.

It should be stressed, however, that although the Casimir effect cannot fully stabilise the wormhole, this does not preclude some very interesting behavior. In particular, it seems  that Casimir effect can slow down the collapse of the wormhole, so that it can have an arbitrarily long lifetime, and moreover, that the collapse may be slow enough to allow a light pulse to traverse the throat before closing. We conclude the paper with an analysis of this phenomenon.

\subsection{Slow Collapse}\label{slowcol}
To understand the collapse of the wormhole, let us begin with the static wormhole spacetime considered in the introduction (\ref{2par}) and promote the radius $a$ to a time dependent quantity $a(t)$:
\begin{align}\label{gcol}\bs
\ud s^2 &= - \ud t^2 + \ud z^2 + A^2\left( \ud\theta ^2 + \sin^2\theta \ud \phi^2 \right),
\\ A&\equiv \sqrt{L^2 + z^2} - L + a(t).\es
\end{align}
The Einstein tensor of this spacetime is then
\begin{align}\nonumber
G_{\muh\nuh} &= \frac{L^2}{(L^2 +z^2)A^2}\mathrm{diag}\left(1,-1,\frac{A}{\sqrt{L^2+z^2}},\frac{A}{\sqrt{L^2+z^2}}\right)\\\nonumber &\quad \,{}+\frac{\dot{a}^2}{A^2}\mathrm{diag}(1,-1,0,0)\\
&\quad \,{} + \mathrm{diag}\left(\frac{-2L^2}{A(L^2 +z^2)^{3/2}},\frac{-2\ddot{a}}{A},\frac{-\ddot{a}}{A},\frac{-\ddot{a}}{A}\right).
\end{align}
Taking the large $L$ limit (or equivalently, focussing on the centre of the throat: $z=0$) we find that
\begin{align}\nonumber
G_{\muh\nuh} &= \frac{1}{a^2}\mathrm{diag}\left(1,-1,\frac{a}{L},\frac{a}{L}\right)+\frac{\dot{a}^2}{a^2}\mathrm{diag}(1,-1,0,0)\\\label{Glongt}
&\quad \,{} + \mathrm{diag}\left(\frac{-2}{La},\frac{-2\ddot{a}}{a},\frac{-\ddot{a}}{a},\frac{-\ddot{a}}{a}\right).
\end{align}

To calculate the Casimir energy-momentum tensor generated by this nonstatic spacetime, one can proceed in a similar fashion to the static case, representing the $L\to\infty$ limit of (\ref{gcol}) as an infinite throat 
\begin{align}\label{Throatt}
\ud s^2 &= - \ud t^2 + \ud z^2 + a^2\left( \ud\theta ^2 + \sin^2\theta \ud \phi^2 \right),
\end{align}
with a time-dependent radius $a$. As this spacetime is independent of $L$, the Casimir energy-momentum must be independent of $L$ also, and so we can perform the following taylor expansion in $\{\dot{a},\ddot{a},\dddot{a},\ldots\}$ about the static case:
\begin{align}\nonumber
T^\mathrm{ren}_{\mu\nu} & = T^\mathrm{ren}_{\mu\nu}[a,\dot{a},\ddot{a},\dddot{a},\ldots],\\\nonumber
& = T^\mathrm{ren}_{\mu\nu}[a,0,0,0,\ldots] \\\label{slowT}
&\quad\, {}\times (1+ O(\dot{a}) + O(\ddot{a}a) + O(\dddot{a}a^2)+ \ldots ),
\end{align}
where the factors of $a$ in each term follow from dimensional analysis. Thus, if the expansion/collapse of the wormhole is sufficiently slow, that is
\begin{align}\label{slow}
|\dot{a}|,|\ddot{a}a|,|\dddot{a}a^2|,\ldots \ll 1,
\end{align}
then we can neglect the extra terms and use the result (\ref{Tren}) that was derived for the static case. Once we have determined the behaviour of $a(t)$, we will be able to confirm the validity of (\ref{slow}); for now, let us take it as given, and neglect the time-derivative terms in equation (\ref{slowT}). Hence $T^\mathrm{ren}_{\mu\nu}$ is simply given by equation (\ref{Tren}) with the radius $a$ now time dependent; let us write this as
\begin{align}\label{Trencol}
T^\mathrm{ren}_{\muh\nuh} &= \frac{\ln(a/a_0)}{1440\pi^2 a^4}\mathrm{diag}(-1,1,\beta-1,\beta-1),
\end{align}
where
\begin{align}
\beta&\equiv \frac{1}{2\ln(a/a_0)}.
\end{align}
Let us also introduce some ordinary matter to the wormhole, with the following energy-momentum tensor:
\begin{align}\nonumber
\kappa &T^\mathrm{ord}_{\muh\nuh}= \left(\frac{\kappa \ln(a/a_0)}{1440\pi^2 a^4} - \frac{2}{La}\right) \mathrm{diag}\left(1,-1,1-\beta,1-\beta \right)\\\nonumber
&\quad\ \ {} +\frac{1}{a^2}\mathrm{diag}\left(1,-1,\frac{a}{L} + \ddot{a}a(2\beta -3),\frac{a}{L} + \ddot{a}a(2\beta -3)\right)\\\label{Tordcol}
&\quad\ \ {} +\frac{\dot{a}^2}{a^2}\mathrm{diag}(1,-1,0,0).
\end{align}
The precise form of this tensor has been chosen to simplify our analysis; for the purposes of this discussion, we need only check that it obeys all the energy conditions, but let us postpone this task until we have found the form of $a(t)$.\footnote{One should also check that the tensor obeys the conservation law $\nabla^\mu T^\mathrm{ord}_{\mu\nu}=0$, where $\nabla_\mu$ is the covariant derivative of the time-dependent throat metric (\ref{Throatt}). Fortunately, this follows from the field equations (\ref{G=T+T}) and the identities $\nabla^\mu T^\mathrm{ren}_{\mu\nu}=0$ and $\nabla^\mu G_{\mu\nu}=0$, which one can check hold true without restriction on $a(t)$.}

Substituting (\ref{Glongt}), (\ref{Trencol}) and (\ref{Tordcol}) into the Einstein field equations (\ref{G=T+T}) and simplifying, we arrive at the following:
\begin{align}
\frac{2\left(1+L\ddot{a}\right)}{La}\mathrm{diag}\left(0,1,\beta-1,\beta-1\right)=0,
\end{align}
the solution of which is of course
\begin{align}\label{add}
\ddot{a} = - 1/L.
\end{align}
Hence the throat radius accelerates towards closure at a constant rate, and fixing $t=0$ to be the time at which $a$ achieves its maximal value $a_{\mathrm{max}}$ we conclude that
\begin{align}\label{a(t)}
a(t)= a_{\mathrm{max}} - \frac{t^2}{2L}.
\end{align}

We are now in a position to confirm the validity of our earlier assumptions. Excluding the unphysical times
\begin{align}
|t|> \sqrt{2L a_{\mathrm{max}}},
\end{align}
for which $a<0$, we see that
\begin{align}
|\dot{a}|= |t|/L \le \sqrt{2a_{\mathrm{max}}/L}\ll 1,
\end{align}
as $L\gg a$. Furthermore, it follows from (\ref{add}) that $|\ddot{a}a|=a/L\ll 1$ and that all higher time-derivatives vanish. This validates the conditions for slow collapse (\ref{slow}).

\begin{figure}
\includegraphics[scale=.45]{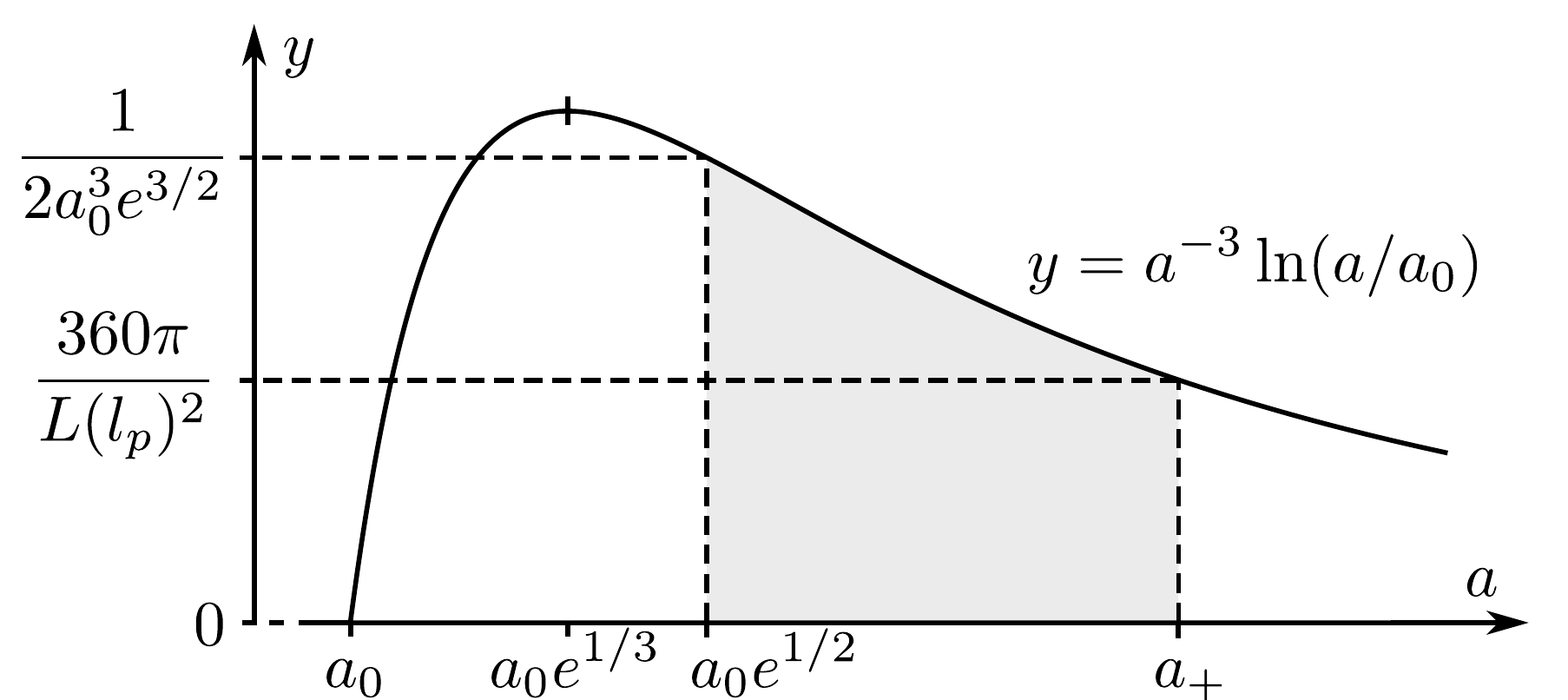}
\caption{Plot of $y=a^{-3}\log(a/a_0)$, indicating the region  $a_0 e^{1/2}\le a\le a_{+}$ in which both inequalities (\ref{2ineq}) are satisfied. It is clear from the graph that $a_{+}$ exists and is greater than 
$a_0 e^{1/2}$ if and only if $360\pi/L(l_p)^2\le 1/2 a_0^3 e^{3/2}$; this condition can be written as (\ref{Lineq}).}\label{logplot}
\centering
\end{figure}

To prove that $T^\mathrm{ord}_{\mu\nu}$ obeys the energy conditions, we shall consider each term of (\ref{Tordcol}) individually. Referring to footnote \ref{nonexotic}, it is clear that the first term on the right-hand-side of (\ref{Tordcol}) will obey all four energy conditions if and only if
\begin{align}
0\le \beta &\le 1&&\text{and}&\frac{\kappa \ln(a/a_0)}{1440\pi^2 a^4}&\ge \frac{2}{La},
\end{align}
or equivalently,
\begin{align}\label{2ineq}
a &\ge a_0 e^{1/2}&&\text{and}& a^{-3}\ln(a/a_0) &\ge \frac{360 \pi}{L (l_p)^2}.
\end{align}
As can be seen from figure \ref{logplot}, both these inequalities can be satisfied if and only if
\begin{align}\label{Lineq}
L\ge a_0 \left(\frac{a_0}{l_p}\right)^2 720 \pi e^{3/2},
\end{align}
in which case they are equivalent to
\begin{align}\label{aineq}
a_0 e^{1/2} \le a \le a_+,
\end{align}
where $a=a_+$ is the larger solution of
\begin{align}\label{a+def}
a^{-3}\ln(a/a_0) &= \frac{360 \pi}{L (l_p)^2},
\end{align}
and can thus be expressed in closed form using the Lambert $W$ function \cite{Corless96}:
\begin{align}\label{W}
a_{+}&= \left[\frac{ - L(l_p)^2}{1080 \pi} W_{-1}\left(\frac{-1080 \pi a_0^3}{L (l_p)^2}\right)\right]^{1/3}.
\end{align}
Note that both conditions (\ref{Lineq}) and (\ref{aineq}) are consistent with the long-throat assumption $L\gg a$, and $a>a_0 > l_p$. Furthermore, we have
\begin{align}\nonumber
a_{+}& = \left(\frac{L (l_p)^2}{360\pi} \ln (a_+/a_0) \right)^{1/3}\ge \left(\frac{L (l_p)^2}{720\pi}\right)^{1/3},
\end{align}
by virtue of (\ref{a+def}) and then (\ref{aineq}); hence the wormhole can be much larger than the Planck scale: $a_{+}\gg l_{p}$ as $L\to \infty$. 

With (\ref{Lineq}) and (\ref{aineq}) taken as given, the other two terms in  $T^\mathrm{ord}_{\mu\nu}$ are essentially trivial. With (\ref{add}) the second term on the right of (\ref{Tordcol}) becomes
\begin{align}\label{Tord2}
\frac{1}{a^2}\mathrm{diag}\left(1,-1,\frac{2a}{L}(2-\beta),\frac{2a}{L}(2-\beta)\right);
\end{align}
recalling that $a/L \ll 1$, and that $0\le \beta \le 1$ as a consequence of (\ref{aineq}), we refer to footnote \ref{nonexotic} to verify that (\ref{Tord2}) obeys all the energy conditions. The last term in (\ref{Tordcol}) also satisfies the specifications of footnote \ref{nonexotic}, and so is similarly nonexotic.

We have thus justified our earlier claims: (i) that the collapse of the wormhole is so slow (\ref{slow}) that one can neglect the time-derivative terms from the renormalised Casimir energy-momentum tensor $T^\mathrm{ren}_{\mu\nu}$, and (ii) that the Einstein equations (\ref{G=T+T}) have been solved by adding only \emph{ordinary} matter, $T^\mathrm{ord}_{\mu\nu}$. As long as $L$ is sufficiently large (\ref{Lineq}) the throat of the wormhole will therefore collapse according to (\ref{a(t)}) with a constant acceleration $\ddot{a}=-1/L$ that can be arbitrarily small.

This picture breaks down if ever the throat radius exits the range (\ref{aineq}), as $T^\mathrm{ord}_{\mu\nu}$ is then required to be exotic, at least for the particular form (\ref{Tordcol}) considered here. If we consider only wormholes for which
\begin{align}\label{amaxineq}
a_{\mathrm{\max}}\le a_{+},
\end{align}
then we need not worry about exceeding the top limit of (\ref{aineq}). However, as the wormhole collapses there will inevitably be a time $t=t_{\mathrm{close}}$ when the bottom limit of (\ref{aineq}) is reached,
\begin{align}\label{amin}
a(t_{\mathrm{close}})=a_0 e^{1/2}\equiv a_{\mathrm{min}},
\end{align}
after which the slow collapse can no longer be supported by the Casimir effect and ordinary matter. At this point, the collapse  presumably becomes much more rapid, and within a very short time the throat radius falls to zero, splitting the wormhole (\ref{gcol}) into two disconnected spacetimes.\footnote{More accurately, the throat radius approaches the Planck scale, wherein quantum effects allow for the change in spacetime topology.} Treating this process as instantaneous in comparison to period of slow collapse, one can easily solve (\ref{a(t)}) and (\ref{amin}) to obtain the closure time for the throat:
\begin{align}
t_\mathrm{close}=\sqrt{2L\left( a_{\mathrm{max}}-a_{\mathrm{min}}\right)}.
\end{align}
As $L$ has no upper bound, we conclude that a long wormhole can exist intact for an arbitrarily long time. Indeed, if we consider the optimal case, where the inequality (\ref{amaxineq}) is saturated, we can use the asymptotic behavior of the Lambert $W$ function,
\begin{align}
W_{-1}(x)\sim \ln(-x) \qquad \text{as}\ \ x\to 0^{-},
\end{align}
to obtain
\begin{align}\label{tsimL}
t_\mathrm{close}&\sim (L^2 l_p)^{1/3}\left(\frac{1}{135\pi}\ln \left(\frac{L(l_p)^2}{1080 \pi a_0^3}\right)\right)^{1/6},
\end{align}
as $L\to \infty$. Thus, under optimal conditions, the closure time grows as $L^{2/3}(\ln{L})^{1/6}$.

Considering that the Casimir effect failed to fully stabilise the wormhole, it is exciting to find that it nonetheless allows the wormhole a remarkable longevity. Indeed, this unexpectedly long lifetime poses an interesting question: can the wormhole remain open long enough for a pulse of light to travel through, thus permitting the transmission of information across the throat? To answer this question definitively lies beyond the scope of this article; however, we shall attempt a preliminary analysis of transmission near the centre of the wormhole.

\subsection{Transmission of Information}\label{trans}
The possibility of light traversing the closing throat cannot be dismissed \emph{a priori}: the wormhole avoids the topological censorship theorem \cite{Friedman93} because the renormalised Casimir energy-momentum tensor $T^\mathrm{ren}_{\mu\nu}$ violates the averaged null energy condition, albeit in a manner that does not allow for absolute stability. Likewise, one cannot immediately conclude that because $t_\mathrm{close}\sim O(L^{2/3}(\ln{L})^{1/6})$ grows more slowly than the throat crossing time $O(L)$, so must transmission fail in the large $L$ limit. The flaw in this reasoning lies in the fact that $t_\mathrm{close}$ is only the time taken for the wormhole to close at $z=0$: we must also account for the propagation of closure outwards from the centre.

\begin{figure}
\includegraphics[scale=.45]{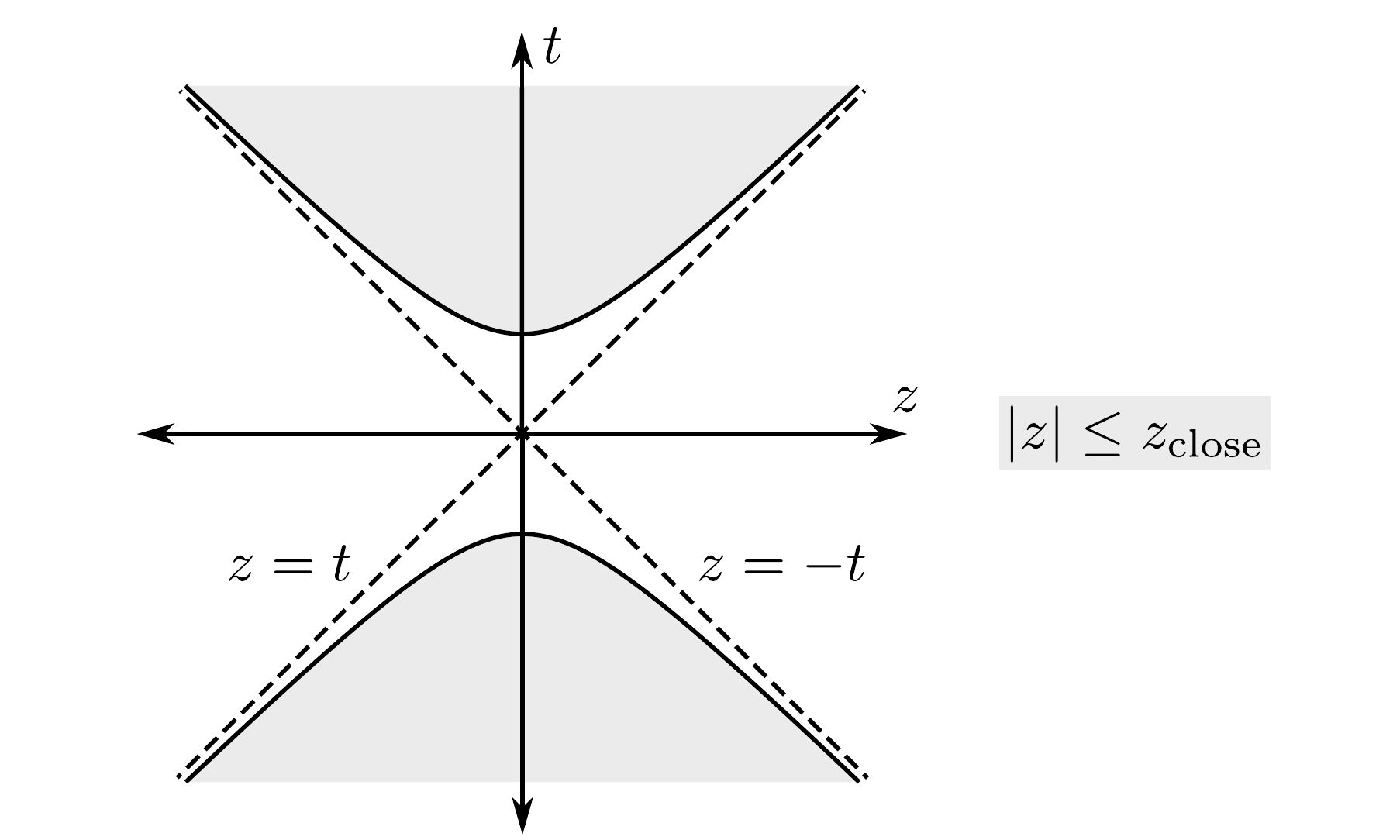}
\caption{Spacetime plot (with angular directions suppressed) illustrating that the null geodesics given in (\ref{nullgeo}) avoid the regions $|z| \le z_\mathrm{close}$ where the throat has closed.}\label{pulse}
\centering
\end{figure}

To proceed we therefore need to consider the approximately quadratic profile of the wormhole (\ref{gcol}) near $z=0$: 
\begin{align}\label{quadprof}
A= a + \frac{z^2}{2L} + O\left(z^4/L^3\right).
\end{align}
Note that in equation (\ref{add}) the acceleration $\ddot{a}$ is independent of the throat radius $a$; thus we do not expect the acceleration to change as the throat widens out, at least to a first approximation for $|z|\ll L$. We therefore model the collapse as \emph{uniform} in close vicinity to the center, with the time-dependent profile obtained simply by inserting (\ref{a(t)}) into (\ref{quadprof}):
\begin{align}
A = a_{\mathrm{max}} + \frac{z^2}{2L} - \frac{t^2}{2L} + O\left(z^4/L^3\right).
\end{align}
Discarding the negligible terms, it follows that the wormhole throat is open everywhere for $|t|< t_\mathrm{close}$, and that when $|t|\ge t_\mathrm{close}$ the wormhole closes ($A =a_\mathrm{min}$) at $z=\pm z_\mathrm{close}(t)$, where
\begin{align}
z_\mathrm{close} = \sqrt{t^2 - 2 L (a_{\mathrm{max}}-a_{\mathrm{min}})}<| t|.
\end{align}
Thus, as illustrated in figure \ref{pulse}, the null geodesics 
\begin{align}\label{nullgeo}
z&=\pm t,&\theta&=\text{const.},&\phi&=\text{const.},
\end{align}
lie entirely within the allowed region $|z|> z_\mathrm{close}$, and hence safely thread the collapsing throat. We conclude that, in the vicinity of the centre of the throat, the wormhole collapses slowly enough to let a pulse of light pass through.

Extending this idea very slightly, we can consider the timelike geodesics
\begin{align}\label{timegeo}
z&=\pm t(1-\epsilon),&\theta&=\text{const.},&\phi&=\text{const.},
\end{align}
with $0<\epsilon\ll 1$, and observe that these wordlines remain within the allowed region out to a distance of $|z|\approx \sqrt{L (a_{\mathrm{max}}-a_{\mathrm{min}})/\epsilon}$. A massive particle, moving sufficiently close to the speed of light ($\epsilon\sim O(a/L)$), can therefore safely traverse the regime of validity of the present analysis $|z|\ll L$.

These are intriguing results: although the wormhole is not \emph{stable}, it seems that it may be \emph{traversable} for particles moving at (or near) the speed of light. Depending on how the two ends of the wormhole connect to the external spacetime, this could allow for communications that exceed the speed of light (from the point of view of external observers) or even closed causal curves. However, it must be stressed that we do not currently know the behaviour of the spacetime far from the centre, so it remains to be seen whether the light pulse can actually escape the wormhole completely. To tackle this problem properly would require, firstly, an extension of the Casimir energy-momentum (\ref{Tren}) beyond infinite throat approximation (\ref{Throatt}), and secondly, a more general model of collapse than the metric (\ref{gcol}). We leave these developments for future work.

\section{Conclusion}
We have obtained the renormalised energy-momentum tensor (\ref{Tren}) induced in the vacuum state of the massless conformally-coupled scalar field by the geometry and topology of a long wormhole throat (\ref{throat}). The tensor describes highly exotic matter (\ref{exotic}) and is sufficiently large that it could, in principle, stabilise a macroscopic wormhole (\ref{a^2}). Unfortunately, the energy-density vanishes along null vectors running parallel to the throat, and this prevents the Casimir effect from stabilising the wormhole in this particular case (\S\ref{ECs}). Nonetheless, the exotic matter provides partial support to the wormhole, allowing it to collapse extremely slowly (\S\ref{slowcol}) and remain open for an arbitrarily long time (\ref{tsimL}). Moreover, near the centre of the throat, the collapse is sufficiently slow that a pulse of light can be safely transmitted (\S\ref{trans}), although it is currently unknown whether this light-pulse can escape the wormhole completely.

These results tentatively suggest that a macroscopic traversable wormhole might be sustained by its own Casimir energy, providing a mechanism for faster-than-light communication and closed causal curves. 
To obtain a more definitive assessment of this possibility, the present research suggests two main avenues of investigation for future work. Firstly, we should explore other traversable wormhole metrics (e.g.\ (\ref{2par}) in the short-throat limit: $L\ll a$) with the hope of finding a spacetime which avoids the throat-parallel null-vector energy-density problem, and can therefore be fully stabilised by its own Casimir energy. Secondly, we should seek to better understand the dynamics of the long-throated wormhole, with a view to establishing whether Casimir-supported collapse does indeed allow information to be transmitted from one end of the wormhole to the other. If either  approach succeeds, it would then be important to determine whether the solutions can survive the introduction of symmetry-breaking perturbations.

\begin{acknowledgments}
The author is supported by a research fellowship at Jesus College, Cambridge, and wishes to also thank Anthony Lasenby and Mike Hobson for helpful advice.
\end{acknowledgments}

\appendix
\section{Evaluation of $I$, $J$ and $K$}\label{Calc}
Let us begin by focussing on $I$ with $\alpha\ge 0$ and $\alphabar\ge 0$. The definitions we will need are
\begin{align}\label{Idef}\bs
I& \equiv \int^\infty_0\! \ud u \left[ \sum_{l=0}^\infty(l+\half) v f(\tfrac{v}{\lambda}) - \int^\infty_0 \!\ud l \, l \vb f(\tfrac{\vb}{\lambda})\right],\\
v&\equiv \sqrt{u^2 +(l+1/2)^2 + \alpha},\quad\vb\equiv \sqrt{u^2 +l^2+\alphabar}.\es
\end{align}
Using the (half-integer) Abel-Plana formula \citep[\S2.2]{Bordag},
\begin{align}
\sum^{\infty}_{l=0}F(l+\half)-\int^{\infty}_{0}\! \ud l F(l)= i \int_0^\infty \ud t \frac{F(-it)- F(it)}{e^{2\pi t} +1},
\end{align}
the bracketed quantity in (\ref{Idef}) becomes
\begin{align}\nonumber
&\sum_{l=0}^\infty(l+\half) v f(\tfrac{v}{\lambda}) - \int^\infty_0 \!\ud l \, l \vb f(\tfrac{\vb}{\lambda}) \\\nonumber
&=i \int^\infty_0\!\!\! \frac{\ud t}{e^{2\pi t} +1}\bigg[(-it) \sqrt{u^2 + (-it)^2 + \alpha}f\\\nonumber
&\quad \,{}- (it) \sqrt{u^2 + (it)^2+ \alpha}f \bigg] + \int_0^\infty \ud l\, l\sqrt{u^2 + l^2 +\alpha}f\\\label{branchcare}
&\quad \,{}- \int_0^\infty \ud l\, l\sqrt{u^2 + l^2+\alphabar}f,
\end{align} 
where, in each case, the argument of $f$ is the adjacent square-root, divided by $\lambda$. Care must be taken with the first two terms in (\ref{branchcare}) as the square-roots must be evaluated by analytically continuing $\sqrt{u^2 +l^2+\alpha}$ from $l\ge 0$ to $l=\pm i t$, $t\ge0$. Following this stipulation, one finds that
\begin{align}
\sqrt{u^2 + (it)^2 +\alpha} = 
 \begin{cases}
   \sqrt{u^2 + (-it)^2 +\alpha} & t^2\le u^2 +\alpha\\
   -\sqrt{u^2 + (-it)^2 +\alpha} &  t^2> u^2 +\alpha 
  \end{cases}.
\end{align}
Furthermore, let us assume that the cut-off function $f(x)$ is an analytic function of $x^2$, so that it takes the same value in both the first and second terms of (\ref{branchcare}), regardless of the value of $t$. Hence
\begin{align}\nonumber
&\sum_{l=0}^\infty(l+\half) v f(\tfrac{v}{\lambda}) - \int^\infty_0 \!\ud l \, l \vb f(\tfrac{\vb}{\lambda}) \\\nonumber
&=\int^{\sqrt{u^2 +\alpha}}_0\! \!\ud t\frac{ 2t \sqrt{u^2 -t^2 + \alpha}}{e^{2\pi t} +1}f+ \int_0^\infty \ud l\, l\sqrt{u^2 + l^2 +\alpha}f\\
&\quad \, {}- \int_0^\infty \ud l\, l\sqrt{u^2 + l^2+\alphabar}f,
\end{align} 
and consequently, $I$ can be split into three pieces: 
\begin{align}\label{3pieces}
I= I_\alpha^{(1)} + I_\alpha^{(2)} -  I_\alphabar^{(2)},
\end{align}
where
\begin{align}\bs\label{I12def}
I^{(1)}_{\alpha} &\equiv  \int^\infty_{0} \!\ud u  \int^{\sqrt{u^2 + \alpha}}_{0}\!\! \ud t \frac{2t \sqrt{u^2 -t^2+ \alpha}}{e^{2\pi t} +1}f,\\
I^{(2)}_{\alpha}&\equiv \int^\infty_{0} \!\ud u \int_{0}^{\infty}\!\ud l\, l \sqrt{u^2 +l^2 + \alpha} f.\es
\end{align}

Concentrating on $I_\alpha^{(1)}$ to begin with, we first swap the order of integration, and then substitute $u^2 = x^2+t^2 -\alpha$, giving
\begin{align}\nonumber
I_\alpha^{(1)}&=\left(\int^{\sqrt{\alpha}}_{0} \!\ud t\, \int^\infty_{0} \!\ud u +  \int^\infty_{\sqrt{\alpha}} \!\ud t \int^\infty_{\sqrt{t^2- \alpha}} \!\ud u \right)\times \\\nonumber
&\qquad\qquad\quad\qquad\qquad\ \ \frac{2 t}{ e^{2\pi t} +1} \sqrt{u^2 -t^2+ \alpha} f
\\ \nonumber
&=\left(\int^{\sqrt{\alpha}}_{0} \!\ud t \int^\infty_{\sqrt{\alpha-t^2}} \!\ud x +  \int^\infty_{\sqrt{\alpha}} \!\ud t\, \int^\infty_{0} \!\ud x \right)\times\\
&\qquad\qquad\qquad\ \ \, \frac{2 t}{ e^{2\pi t} +1} \frac{x^2}{\sqrt{x^2 +t^2- \alpha}} f(x/ \lambda) .
\end{align}
Notice that the assumption $\alpha \ge 0$ was necessary at this step.

At this point, we shall take $f$ to be a sharp cut-off (i.e.\ $f(x)\equiv H(1 - x^2)$, $H$ being the Heaviside step function)\footnote{We have written $H(1-x^2)$ rather than $H(1-x)$ to be consistent with our previous specification that $f(x)$ be an analytic function of $x^2$. Of course, the Heaviside step function is not analytic, but it can be thought of as the limit of an analytic sigmoid function as the width of its step is taken to zero.} and perform the $x$-integration using hyperbolic substitutions:
\begin{align}\nonumber
I_\alpha^{(1)}&= \int^{\sqrt{\alpha}}_{0} \!\ud t \frac{2 t}{ e^{2\pi t} +1}\, \int^\lambda_{\sqrt{\alpha-t^2}} \!\ud x \frac{x^2}{\sqrt{x^2 +t^2- \alpha}}
\\\nonumber
&\quad \, {} + \int^\infty_{\sqrt{\alpha}} \!\ud t\frac{2 t}{ e^{2\pi t} +1} \int^\lambda_{0} \!\ud x \frac{x^2}{\sqrt{x^2 +t^2- \alpha}}
\\\nonumber
&= \int^{\sqrt{\alpha}}_{0} \!\ud t\,\frac{2t (\alpha - t^2)}{ e^{2\pi t} +1} \int^{\mathrm{arcosh}(\lambda/\sqrt{\alpha-t^2})}_{0}\!\ud y\,\cosh^2\!y 
\\\nonumber
&\quad\, + \int^{\infty}_{\sqrt{\alpha}} \!\ud t\,\frac{2t (t^2-\alpha)}{ e^{2\pi t} +1} \int^{\mathrm{arsinh}(\lambda/\sqrt{t^2 -\alpha})}_{0}\!\ud y\,\sinh^2\!y
\\\nonumber
&= \int^{\infty}_{0} \!\ud t\,\frac{t (\alpha - t^2)}{ e^{2\pi t} +1} \bigg[ \ln\left(\lambda + \sqrt{\lambda^2 +t^2-\alpha} \right)
\\
&\qquad\ \ - \frac{1}{2}\ln |t^2-\alpha|+ \frac{\lambda}{\alpha-t^2}\sqrt{\lambda^2+ t^2-\alpha}\bigg].
\end{align}
We are now in a position to expand the integrand as a taylor series in $\lambda^{-1}$:
\begin{align}\nonumber
I_\alpha^{(1)}&=  \int^{\sqrt{\lambda^2 +\alpha}}_{0} \!\ud t\,\frac{t (\alpha - t^2)}{ e^{2\pi t} +1} \bigg[ \ln\left(2\lambda+ O\left(\lambda^{-1}\right)\right) \\\label{logtaylor}
&\qquad-\frac{1}{2}\ln|t^2-\alpha| + \frac{\lambda^2}{\alpha - t^2}  - \frac{1}{2} +O\left(\lambda^{-2}\right) \bigg] + P,
\end{align}
where
\begin{align}\nonumber
P&\equiv  \int^{\infty}_{\sqrt{\lambda^2 +\alpha}} \!\ud t\,\frac{t (\alpha - t^2)}{ e^{2\pi t} +1} \bigg[ \ln\left(\lambda + \sqrt{\lambda^2 +t^2-\alpha} \right)  \\
&\qquad\ \ -\frac{1}{2} \ln |t^2-\alpha|+ \frac{\lambda}{\alpha-t^2}\sqrt{\lambda^2+ t^2-\alpha}\bigg]
\end{align}
is the part of the integral where $\lambda^2< |t^2-\alpha|$ and so the expansion cannot be performed. Fortunately,  $P$ is exponentially suppressed as $\lambda\to\infty$, and so can safely be neglected:
\begin{align}
P=O\left(\int^\infty_\lambda \!\ud t\, t^3 e^{-2\pi t}\ln t\right)= O\left(e^{-2\pi \lambda}\lambda^3 \ln \lambda \right).
\end{align}
Hence, equation (\ref{logtaylor}) becomes
\begin{align}\nonumber
I_\alpha^{(1)}&= \int^{\infty}_{0} \!\ud t\,\frac{t}{ e^{2\pi t} +1} \bigg[\lambda^2 + (\alpha-t^2)\ln (2\lambda) \\
&\quad\,{} + \frac{ t^2- \alpha}{2}\left(1+ \ln|t^2 -\alpha| \right) \bigg]- Q+ O\left(\lambda^{-2}\right),
\end{align}
where
\begin{align}\nonumber
Q&\equiv   \int^{\infty}_{\sqrt{\lambda^2 + \alpha}}  \!\ud t\,\frac{t}{ e^{2\pi t} +1} \bigg[\lambda^2 + (\alpha-t^2)\ln (2\lambda) \\\nonumber
&\qquad\qquad + \frac{ t^2- \alpha}{2}\left(1+ \ln|t^2 -\alpha|  \right) \bigg]\\
&= O\left(e^{-2\pi \lambda}\lambda^3 \ln \lambda \right)
\end{align}
is also negligible.

Using the standard results
\begin{align}\bs
\int^{\infty}_{0}\!\ud t \, \frac{t}{e^{2\pi t} +1}&= \frac{1}{48},\\ \int^{\infty}_{0}\!\ud t \, \frac{t^3}{e^{2\pi t} +1}&=\frac{7}{1920}, \es
\end{align}
and defining
\begin{align}\label{Xdef}
X(\alpha)\equiv\frac{1}{2} \int^{\infty}_{0}\!\ud t  \, \frac{t(t^2-\alpha)\ln |t^2 -\alpha|}{e^{2\pi t} +1},
\end{align}
we conclude that
\begin{align}\nonumber
I^{(1)}_\alpha &= \frac{1}{48}\left(\lambda^2 + \left( \alpha - \frac{7}{40} \right)\left(\ln(2\lambda) - \frac{1}{2} \right)\right) + X(\alpha)\\\label{I1}
&\quad\, + O(\lambda^{-2}).
\end{align}

All that remains is to calculate the integral $I^{(2)}_\alpha$ that appears in equation (\ref{3pieces}). Swapping the order of integration, substituting $u^2 = x^2 - l^2 -\alpha$, and taking $f$ to be a sharp cut-off, we have
\begin{align}\nonumber
I^{(2)}_{\alpha}&= \int^{\infty}_{0} \!\ud l  \int^{\infty}_{0} \!\ud u \,  l \sqrt{u^2 + l^2 + \alpha} f \\\nonumber
&= \int^{\infty}_{0} \!\ud l  \int^{\infty}_{\sqrt{l^2 +\alpha}} \!\ud x   \frac{l x^2  f(x/\lambda)}{\sqrt{x^2 - l^2 - \alpha} } \\
&= \int^{\sqrt{\lambda^2 -\alpha}}_{0} \!\ud l  \int^{\lambda}_{\sqrt{l^2 +\alpha}} \!\ud x  \frac{l x^2}{\sqrt{x^2 - l^2 - \alpha}}.
\end{align}
Swapping the order of integration once more,  
\begin{align}\nonumber
I^{(2)}_{\alpha}&=\int^{\lambda}_{\sqrt{\alpha}} \!\ud x \, x^2  \int^{\sqrt{x^2- \alpha}}_{0} \!\ud l  \frac{ l}{\sqrt{x^2 - l^2 - \alpha}}\\\nonumber
&=  \int^{\lambda}_{\sqrt{\alpha}} \!\ud x \, x^2  \sqrt{x^2 - \alpha}\\\nonumber
&=  -\frac{\alpha^2}{8} \left(\ln\left(\lambda + \sqrt{\lambda^2-\alpha}\right) - \frac{1}{2}\ln\alpha\right) \\
&\quad\, + \frac{\lambda}{8} (2\lambda^2 - \alpha)\sqrt{\lambda^2 -\alpha}.
\end{align}
Thus, as $\lambda\to\infty$
\begin{align}\nonumber
I^{(2)}_{\alpha}&=\frac{\lambda^4}{4}-\frac{\alpha }{4}\lambda^2 - \frac{ \alpha^2 }{8}\ln(2\lambda) + \frac{\alpha^2 }{32} \left( 2 \ln\alpha+1\right)\\&\quad\, + O (\lambda^{-2}),
\end{align}
and
\begin{align}\nonumber
I^{(2)}_{\alpha}- I^{(2)}_{\alphabar}&=\frac{\alphabar-\alpha}{4}\lambda^2 + \frac{\alphabar^2- \alpha^2 }{8}\ln(2\lambda)\\\nonumber
&\quad\, + \frac{\alpha^2 }{32} \left(2 \ln\alpha+1\right) -\frac{\alphabar^2 }{32} \left(2 \ln\alphabar+1\right)\\\label{I2}&\quad\, + O (\lambda^{-2}).
\end{align}

We now have all the results we need. Inserting (\ref{I1}) and (\ref{I2}) into equation (\ref{3pieces}), we arrive at
\begin{align}\nonumber
I&=\\\nonumber
&{}\frac{1+12(\alphabar-\alpha)}{48} \lambda^2+ \frac{1}{48}\left( \alpha + 6(\alphabar^2- \alpha^2) -\frac{7}{40}\right)\ln(2\lambda)    \\\nonumber
&{} -\frac{1}{96}\left(\alpha -\frac{7}{40}\right)+ \frac{1}{16}\left(\alpha^2 \ln \alpha - \alphabar^2 \ln \alphabar \right)\\\label{Ipreresult}
&{}+\frac{1}{32}(\alpha^2-\alphabar^2)+X(\alpha) + O (\lambda^{-2}),
\end{align}
which, according to our original assumption, is valid for $\alpha \ge 0$ and $\alphabar \ge 0$. To obtain the result for $\alpha \le 0$, we can analytically continue the above expression, moving $\alpha$ to negative values while avoiding the branch-point $\alpha =0$ in the complex plane. The only subtlety to this step is that the $\ln \alpha$ term in (\ref{Ipreresult}) will produce $\pm i\pi$ in the process, the sign depending on whether one moves clockwise or anticlockwise. Clearly, the same can be said of $\alphabar$. Fortunately, we are only interested in the real part of $I$, so this ambiguity is of no concern:
\begin{align}\nonumber
\Re\{&I\}=\\\nonumber
&{}\frac{1+12(\alphabar-\alpha)}{48} \lambda^2+ \frac{1}{48}\left( \alpha + 6(\alphabar^2- \alpha^2) -\frac{7}{40}\right)\ln(2\lambda)    \\\nonumber
&{} -\frac{1}{96}\left(\alpha -\frac{7}{40}\right) + \frac{1}{16}\left(\alpha^2 \ln |\alpha| - \alphabar^2 \ln |\alphabar| \right)\\\label{Iresult}
& {}+\frac{1}{32}(\alpha^2-\alphabar^2) +X(\alpha) + O (\lambda^{-2}),
\end{align}
valid for all $\alpha,\alphabar \in \mathbb{R}$.

The calculations for $J$ and $K$ proceed in exactly the same fashion, so it serves no purpose to repeat them here. The results (valid for all $\alpha,\alphabar \in \mathbb{R}$) are as follows:
\begin{align}\nonumber
\Re\{&J\}=\\\nonumber
&{}\frac{1+12(\alphabar-\alpha)}{48} \lambda^2- \frac{1}{48}\left( \alpha + 6(\alphabar^2- \alpha^2) -\frac{7}{40}\right)\ln(2\lambda)    \\\nonumber
&{} -\frac{1}{96}\left(\alpha -\frac{7}{40}\right)  - \frac{1}{16}\left(\alpha^2 \ln |\alpha| - \alphabar^2 \ln |\alphabar| \right)\\\label{Jresult}
& {}+\frac{3}{32}(\alpha^2-\alphabar^2)-X(\alpha) + O (\lambda^{-2}),
\end{align}
and
\begin{align}\nonumber
\Re\{K\}&=\frac{1+12(\alphabar-\alpha)}{24} \ln(2\lambda)+\frac{1}{4}\left(\alpha \ln |\alpha| - \alphabar\ln |\alphabar|\right)  \\\label{Kresult}
&\quad \,+\frac{1}{4}(\alphabar - \alpha) 
- 2Y(\alpha)  + O (\lambda^{-2}),
\end{align}
where
\begin{align}\label{Ydef}
Y(\alpha)\equiv\frac{1}{2} \int^{\infty}_{0}\!\ud t  \, \frac{t\ln |t^2 -\alpha|}{e^{2\pi t} +1}.
\end{align}

\section{Energy Inequalities}\label{EnAp}
Here we derive some key inequalities obeyed by the renormalised Casimir energy-momentum tensor (\ref{Tren}) of the long wormhole throat (\ref{throat}). Considering that we are interested in the creation of exotic matter, let us assume
\begin{align}\label{a>a0}
a>a_0,
\end{align}
so that the weak and dominant energy conditions are immediately violated by virtue of $T^\mathrm{ren}_{00}<0$. We can then write
\begin{align}\label{Tprop}
T^\mathrm{ren}_{\muh\nuh}\propto \mathrm{diag}(-1,1,\beta-1,\beta-1),
\end{align}
where we have introduced
\begin{align}
\beta\equiv \frac{1}{2 \ln (a/a_0)},
\end{align}
and use the symbol $\propto$ to indicate proportionality by means of a \emph{positive} constant:
\begin{align}
P_{\mu\nu}&\propto Q_{\mu\nu}& &\Rightarrow&  \exists\ s&>0& &\text{s.\ t.}& P_{\mu\nu}&=sQ_{\mu\nu}.
\end{align}

The various energy conditions concern the energy-density and energy-flux defined by a four-velocity $v^\mu$ that is either timelike or null; without loss of generality, let us take this four-velocity to be
\begin{align}
v^\muh& \propto (1,v_z,v_\perp,0),& v_z^2+v_\perp^2&\le 1.
\end{align}
The energy-density $v^\mu T^\mathrm{ren}_{\mu\nu} v^\nu$ and energy current $T^\mathrm{ren}_{\mu\nu} v^\nu$ then obey
\begin{align}\bs
v^\mu T^\mathrm{ren}_{\mu\nu} v^\nu&\propto -1 + v_z^2+v_\perp^2(\beta-1)^{\phantom{2}} \le v_\perp^2(\beta-2), \\
( T^\mathrm{ren}_{\mu\nu} v^\nu)^2 &\propto -1 + v_z^2 + v_\perp^2 (\beta-1)^2\le v_\perp^2(\beta-2)\beta, \es
\end{align}
with equality if and only if the four-velocity is null. Thus $0<\beta<2$ ensures that the energy-flux is always causal ($T^\mathrm{ren}_{\mu\nu} v^\nu$ is never spacelike) and \emph{every} four-velocity defines a negative energy-density, except for null vectors directly parallel to the throat, which have zero energy-density. That is,
\begin{align}\label{Tineqs}
0<\beta&< 2 &\Rightarrow&& v^\mu T^\mathrm{ren}_{\mu\nu} v^\nu&\le 0,& (T^\mathrm{ren}_{\mu\nu} v^\nu)^2 &\le 0,
\end{align}
with equality if and only if $v_z=\pm1$. Note that we already had $\beta>0$ as a consequence of the assumption (\ref{a>a0}), and that the full restriction $0<\beta<2$ is equivalent to
\begin{align}\label{a>ae1/4}
a > a_0 e^{1/4}.
\end{align}
In contrast, $\beta >2$ preserves the null energy condition
\begin{align}
\beta&> 2,\quad v^\mu v_\mu=0 &\Rightarrow&& v^\mu T^\mathrm{ren}_{\mu\nu} v^\nu&\ge 0,
\end{align}
but allows for noncausal energy-flux:
\begin{align}
\beta&> 2&\Rightarrow&& \exists\ v^\mu\quad \mathrm{s.\ t.}\quad( T^\mathrm{ren}_{\mu\nu} v^\nu)^2&> 0.
\end{align}
If $\beta=2$, then $T^\mathrm{ren}_{\mu\nu}\propto g_{\mu\nu}$ resembles a cosmological constant term, trivially obeying the null energy condition,
\begin{align}
\beta&= 2,\quad v^\mu v_\mu=0 &\Rightarrow&& v^\mu T^\mathrm{ren}_{\mu\nu} v^\nu= 0,
\end{align}
with causal energy-flux:
\begin{align}
\beta&= 2&\Rightarrow&&( T^\mathrm{ren}_{\mu\nu} v^\nu)^2&\le 0.
\end{align}

Finally, to asses the strong energy condition we construct the trace-reverse energy-momentum tensor
\begin{align}\nonumber
\overline{T}^\mathrm{ren}_{\muh\nuh} &\equiv T^\mathrm{ren}_{\muh\nuh}  - \eta_{\mu\nu} T^\mathrm{ren}/2\\
& \propto \mathrm{diag}(\beta-1,1-\beta,-1,-1),
\end{align}
and observe that 
\begin{align}
v^\mu \overline{T}^\mathrm{ren}_{\mu\nu} v^\nu&\propto (\beta-1) + (1-\beta) v_z^2 - v_\perp^2 \ge (\beta-2)(1-v_z^2),
\end{align}
so that if $\beta\ge 2$ the strong energy condition is obeyed. If $0<\beta <2$ then we can consider a timelike four-velocity with $v_z=0$ and $v_\perp = 1- \epsilon (2-\beta)$, where $\epsilon$ in an arbitrarily small positive number; this gives
\begin{align}\nonumber
v^\mu \overline{T}^\mathrm{ren}_{\mu\nu} v^\nu& \propto (\beta-1) - (1- \epsilon (2-\beta))= (\beta -2)(1-\epsilon) \\
&\propto (\beta-2)<0,
\end{align}
in violation of the strong energy condition.

In aggregate, these results demonstrate that the renormalised Casimir energy-momentum tensor (\ref{Tren}) violates all four energy conditions (null, weak, dominant and strong) if and only if (\ref{a>ae1/4}) is obeyed. Under this restriction, the inequalities (\ref{Tineqs}) hold true, and are saturated if and only if $v^\mu \propto (1,\pm1,0,0)$.

\bibliography{long}
\end{document}